\documentclass[11pt]{article}
\usepackage{}
\usepackage{amssymb}
\usepackage{amsfonts}
\usepackage{mathrsfs}
\usepackage{graphicx}
\usepackage{epstopdf}
\usepackage{amsmath}
\allowdisplaybreaks[4]
\usepackage{color}
\usepackage{amsthm}
\usepackage{amsfonts}
\usepackage{graphicx}
\usepackage{epstopdf}
\usepackage{color}
\usepackage{mathrsfs}
\usepackage{enumerate}
\usepackage{tikz}
\usepackage{epstopdf}
\usepackage{pifont,bm}
\usepackage{tikz}
\usepackage{enumerate}

\allowdisplaybreaks[4]

\theoremstyle{definition}

\makeatletter
\def\@biblabel#1{[#1]}
\makeatother

\makeatletter \@addtoreset{equation}{section}

\begin{document}

\begin{titlepage}
\title{\bf{Simple and high-order $N$-solitons of the nonlocal generalized Sasa-Satsuma equation via an improved Riemann-Hilbert method
\footnote{This work is supported
by the National Natural Science Foundation of China (No.12271129 and No.12201622), the China Scholarship Council (No.202206120152), the
China Postdoctoral Science Foundation (No.2023M733404), the Young Innovative Talents Project of Guangdong Province of China (No.2022KQNCX104) and the Guangdong Basic and Applied Basic Research Foundation (No.2022A1515111209).\protect\\
\hspace*{3ex}$^{*}$Corresponding authors.\protect\\
\hspace*{3ex} E-mail addresses: guixianwang@hit.edu.cn (G.X. Wang), xbwang@cumt.edu.cn (X.-B. Wang), haie\_long@smbu.edu.cn (H.E. Long), bohan@hit.edu.cn (B. Han)}
}}
\author{Guixian Wang$^{1}$, Xiu-Bin Wang$^{2}$, Haie Long$^{3}$, Bo Han$^{1,*}$\\
\small \emph{$^{1}$School of Mathematics, Harbin Institute of Technology, Harbin 150001, China} \\
\small \emph{$^{2}$School of Mathematics and Institute of Mathematical Physics,}\\
\small \emph{China University of Mining and Technology, Xuzhou 221116, China} \\
\small \emph{$^{3}$Department of Computational Mathematics and Cybernetics,}\\
\small \emph{Shenzhen MSU-BIT University, 518172, Shenzhen, China}\\
\date{}}
\thispagestyle{empty}
\end{titlepage}
\maketitle

\vspace{-0.5cm}
\begin{center}
\rule{15cm}{1pt}\vspace{0.3cm}

\parbox{15cm}{\small
{\bf Abstract}\\
\hspace{0.5cm}
In this paper, we investigate the nonlocal generalized Sasa-Satsuma (ngSS) equation based on an improved Riemann-Hilbert method (RHM). Different from the traditional RHM, the $t$-part of the Lax pair plays a more important role rather than the $x$-part in analyzing the spectral problems. So we start from the $t$-part of the spectral problems. In the process of dealing with the symmetry reductions, we are surprised to find that the computation is much less than the traditional RHM. We can more easily derive the compact expression of $N$-soliton solution of the ngSS equation under the reflectionless condition. In addition, the general high-order $N$-soliton solution of the ngSS equation is also deduced by means of the perturbed terms and limiting techniques. We not only demonstrate different cases for the dynamics of these solutions in detail in theory, but also exhibit the remarkable features of solitons and breathers graphically by demonstrating their 3D, projection  profiles and wave propagations. Our results should be significant to understand the nonlocal nonlinear phenomena and provide a foundation for fostering more innovative research that advances the theory.
}
\vspace{0.2cm}

\parbox{15cm}{\small{

\vspace{0.3cm} \emph{Key words:} Nonlocal generalized Sasa-Satsuma equation; Improved Riemann-Hilbert method; Inverse scattering transform; Soliton solutions.

 } }
\end{center}
\vspace{0.3cm} \rule{15cm}{1pt} \vspace{0.2cm}

\section{Introduction}
After the publication of Ablowitz-Kaup-Newell-Segur (AKNS) in 1974 \cite{Ablowitz-1974}, it was widely believed that all the fundamental and meaningful symmetry reductions of the classical AKNS scattering problem had already been uncovered. In 2013, however, Ablowitz, et al. unveiled a new reduction within the realm of parity-time symmetry. This reduction resulted in the emergence of a nonlocal nonlinear Schr\"{o}dinger (NLS) equation, which showcased a unique soliton solution. It is truly astonishing that the AKNS symmetry reduction discovered in \cite{Ablowitz-2013} represents merely the initial stage of the journey. These nonlocal symmetry reductions can be in time-only, space-only, and even both in space and time. Each new symmetry condition induce its own new nonlocal nonlinear integrable equation. The investigation of nonlocal integrable equations \cite{Ablowitz-2016,Ablowitz-2018} has become one of the most popular topics in soliton theory and nonlinear mathematical physics.

As a higher-order NLS equation, the Sasa-Satsuma (SS) equation \cite{SS-1991} is an important integrable equation and plays significant roles in explaining the propagation of femtosecond pulses in optical fibers \cite{K-1985,Xu-2013}, soliton propagation in water waves \cite{Xu-2015,J-2015}, plasma physics \cite{X-2014,Zhao-2014,AM-2021}, etc. One of generalized forms of the SS equation \cite{Lv-2009,Geng-2016,Geng-2017} reads
\begin{align}\label{nss}
&u_{t}(x,t)+u_{xxx}(x,t)+6\left[\alpha u(x,t)v(x,t)u_{x}(x,t)+\beta u^{2}(x,t)u_{x}(x,t)
\right]\notag\\
&+3\left[\alpha u(x,t)(u(x,t)v(x,t))_{x}
+\beta^{\ast}v(x,t)(u(x,t)v(x,t))_{x}\right]=0,
\end{align}
and Eq. \eqref{nss} arouses wide attention. In \cite{Zhu-2017}, authors constructed periodic solutions and some localized solutions of the reverse space-time nonlocal SS equation \eqref{nss} with $v(x,t)=u^{\ast}(-x,-t)$ via the binary Darboux transformation (DT) method. In \cite{Chen-2022}, authors obtained the general soliton and high-order soliton solutions for Eq. \eqref{nss} with $v(x,t)=u(-x,-t)$ by solving the Riemann-Hilbert problem (RHP). In \cite{Zhu-2023}, some new properties for Eq. \eqref{nss} with $v(x,t)=u^{\ast}(-x,-t)$ were found by using the DT. Three cases of nonlocal SS equation were explored based on the inverse scattering transforms in \cite{Wang-2022}.

The rising laser power has sparked a growing fascination in the propagation of femtosecond soliton pulses within birefringent or multimode fibers, as well as the phenomenon of pulse trapping that arises from these pulses across the zero-dispersion wavelength. Luckily, the above dynamics can be characterised by the generalized coupled SS equation \cite{Ma-2023} of the form
\begin{align}\label{gcss}
q_{1,t}(x,t)&+q_{1,xxx}(x,t)+6\sigma\left(|q_{1}(x,t)|^2+|q_{2}(x,t)|^2
\right)q_{1,x}(x,t)\notag\\
&+3\sigma q_{1}(x,t)\left(|q_{1}(x,t)|^2+|q_{2}(x,t)|^2
\right)_{x}=0,\notag\\
q_{2,t}(x,t)&+q_{2,xxx}(x,t)+6\sigma\left(|q_{1}(x,t)|^2+|q_{2}(x,t)|^2
\right)q_{2,x}(x,t)
\notag\\
&+3\sigma q_{2}(x,t)\left(|q_{1}(x,t)|^2+|q_{2}(x,t)|^2
\right)_{x}=0,
\end{align}
where $q_{1}$ and $q_{2}$ denote the optical fields, $\sigma$ is the ratio of the width of the spectra to the carrier frequency, the last three terms severally stand for the effects of third-order dispersion, self-steepening and stimulated Raman scattering. Additionally, the long-time behavior of the solutions of Eq. \eqref{gcss} with $\sigma=1$ was analyzed in \cite{XB-2021}.

By making a nonlocal symmetry reduction $q_{2}(x,t)=q_{1}(-x,t)$ for the generalized coupled SS system \eqref{gcss}, it is natural to obtain the nonlocal generalized SS (ngSS) equation
\begin{align}\label{ngss}
q_{t}(x,t)&+q_{xxx}(x,t)+6\sigma\left(|q(x,t)|^2+|q(-x,t)|^2
\right)q_{x}(x,t)\notag\\
&+3\sigma q(x,t)\left(|q(x,t)|^2+|q(-x,t)|^2
\right)_{x}=0.
\end{align}
It describes the solutions of the generalized coupled SS system \eqref{gcss} under the special initial condition $q_{2}(x,0)$ = $q_{1}(-x,0)$. To be more precise, the $q_{1}$ and $q_{2}$ components are related by parity symmetry, which plays a dominant role in the nonlinear wave propagation in such physical circumstances under the nonlocal symmetry reduction of the initial conditions. This physical interpretation can facilitate us comprehend the distinctive solution states induced by the nonlocal symmetry reduction. Consequently, the ngSS equation \eqref{ngss} holds significant physical meaningful in this context. However, there a few researches conducted on the ngSS system. In \cite{XW-2023}, authors only explored the general multi-solitons for the nonlocal reverse-time SS equation under the constraint $q_{2}(x,t)=q_{1}^{\ast}(x,-t)$.

Compared Eq. \eqref{nss} with Eq. \eqref{ngss}, we find that the nonlinearly induced potential $u(x,t)u^{\ast}(-x,-t)$ in Eq. \eqref{nss} is generally complex-valued and symmetric in $x$ and $t$, however, $|q(x,t)|^{2}+ |q(-x,t)|^{2}$ in Eq. \eqref{ngss} is real-valued and symmetric in $x$. On the side, the associated matrix spectral problem for \eqref{nss} is a reduced $3 \times 3$ AKNS spectral problem, while that for \eqref{ngss} is a reduced $5 \times 5$ spectral problem.

For the nonlocal SS equation \eqref{nss}, the symmetry relations of the discrete scattering data were found to be simple soliton, so the multi-soliton solutions were easily calculated in terms of the  Riemann-Hilbert method (RHM) \cite{Chen-2022}. The RHM was first presented by Gardner et al. in 1967 \cite{CS-1967} for the KdV equation. Then the dressing Zakharov-Shabat technique was used to construct the soliton solutions \cite{VE-1974,VE-1979}. It is noted that the spectral analysis usually starts from the $x$-part of the Lax pair in the classical RHM, such as these references \cite{Biondini-2014,Biondini-2016,Yang-2018,Yan-2020,
Ma-2020,Ma-2022,Feng-2021,GX-2023}. Nevertheless, when it comes to \eqref{ngss}, establishing the symmetry relations of the discrete scattering data through the conventional RHM, which involves spectral analysis from the spatial component of the Lax pair, proves to be extremely arduous. Hence the task of deducing symmetry relations for the discrete scattering data and obtaining a general $N$-soliton solution for Eq. \eqref{ngss} is riddled with uncertainties and formidable obstacles.

Inspired by the research work \cite{Wu-2019} and \cite{Wu-2023}. In these two references, the spectral analysis of the Newell-type long-wave-short-wave equation and the nonlocal integrable NLS equation were conducted focusing on the temporal component of the Lax pair. It aims to construct a set of analytical column spectral functions that effectively express the RHP. Therefore, we presume that the symmetry relations of the discrete scattering data and then the general simple multi-soliton solution might be found for Eq. \eqref{ngss}.

It is known to all that the high-order soliton solution plays an important role in characterizing a weak bound state of solitons as well as the study of train propagation of solitons with nearly equal velocities and amplitudes \cite{Gagnon-1994}. To the best of our knowledge, there are many work on high-order solitons of local \cite{Villarroel-1999,Ablowitz-2000,Bian-2015,Bo-2019,Fan-2020,
Tian-2022,Mao-2023} and nonlocal equations \cite{Chen-2022,Bo-2018,Min-2023}. However, high-order multi-solitons for Eq. \eqref{ngss} has never been reported.

Thus, in this paper, we shall study general $N$-soliton and high-order $N$-soliton solutions of the ngSS equation \eqref{ngss} with the help of the improved RHM.

The frame of this paper is arranged as follows. Sec. 2 aims to explain the direct scattering transform of the ngSS equation \eqref{ngss} by analyzing spectral problem from the $t$-part of the Lax pair. We construct two matrix functions and analyze their asymptotic behaviors. Then spacial evolution is considered to introduce the RHP, its symmetry reductions are also analyzed. Sec. 3 gives the compact expression of $N$-soliton solution for Eq. \eqref{ngss} under the reflectionless condition. We illustrate different cases theoretically in terms of the solution. Specially, for the case of $1$-soliton, we give a proposition about amplitudes before and after the collision. Then the numerical results further confirm our analysis. In sec. 4, we derive the $N_{0}$-th high-order $N$-soliton solution in view of the perturbed terms and limiting techniques. Then we deduce the case of the second-order $1$-soliton and the third-order $2$-soliton solutions in detail. What's more, some novel dynamic behaviors are exhibited graphically. Finally, we present our conclusions and engage in discussions.

\section{Direct scattering transform}

The Lax pair for the ngSS equation \eqref{ngss} is written as
\begin{align*}
&Y_{x}=XY,~~X:=X(x,t,k)=\textrm{i}k\sigma_{3}+Q,\notag\\
&Y_{t}=TY,~~T:=T(x,t,k)=4\textrm{i}k^{3}\sigma_{3}+\widetilde{Q},
\end{align*}
where $Y:=Y(x,t,k)$ is a column vector function of the spectral parameter $k$, $q:=q(x,t)$ is a complex-valued function, the usual
matrix commutator is defined as $\left[X,T\right]=XT-TX$, and $\sigma_{3}=\mathrm{diag}\left(1,1,1,1,-1\right)$,
\begin{align*}
&Q:=Q(x,t)=\left(
                         \begin{array}{ccccc}
                           0 & 0 & 0 & 0 & q(x,t) \\
                           0 & 0 & 0 & 0 & \sigma q^{\ast}(x,t) \\
                           0 & 0 & 0 & 0 & q(-x,t) \\
                           0 & 0 & 0 & 0 & \sigma q^{\ast}(-x,t) \\
                           -\sigma q^{\ast}(x,t) & -q(x,t) & -\sigma q^{\ast}(-x,t) & -q(-x,t) & 0 \\
                         \end{array}
                       \right),\notag\\
&\widetilde{Q}:=\widetilde{Q}(x,t)=4k^{2}Q+2\textrm{i}k
\left(Q^{2}+Q_{x}\right)\sigma_{3}+\left[Q_{x},Q\right]
-Q_{xx}+2Q^{3}.
\end{align*}
It is easy to verify that Eq. \eqref{ngss} can be derived by the compatibility condition $X_{t}-T_{x}+\left[X,T\right]=0$.

The above Lax pair is a $5\times 5$ matrix and different from the one of the nonlocal SS equation \eqref{nss}, which leads to a more complicated spectral analysis.

Without loss of generality, we focus on the case where $\sigma=1$ in the following spectral analysis.

\subsection{Spectral analysis with temporal part}

Based on the transformation
\begin{align}\nonumber
\mu:=\mu(x,t,k)=Y\textrm{e}^{-\textrm{i}k\sigma_{3}x-4\textrm{i}^{3}\sigma_{3}t},
\end{align}
we introduce a new matrix $\mu$, which satisfies the following Lax pair
\begin{align}\label{af-3}
&\mu_{x}=\textrm{i}k\left[\sigma_{3},\mu\right]+Q\mu,\notag\\
&\mu_{t}=4\textrm{i}k^{3}\left[\sigma_{3},\mu\right]+\widetilde{Q}\mu.
\end{align}
Then we begin analyze the $t$-part of the Lax pair \eqref{af-3}.

Define $\mu_{\pm}:=\mu(t,k)$ be two solutions of the second equation in \eqref{af-3}, to be more precise,
\begin{align}\label{af-4}
\mu_{\mp}=\left(\left[\mu_{\mp}\right]_{1},\left[\mu_{\mp}\right]_{2}
,\left[\mu_{\mp}\right]_{3},\left[\mu_{\mp}\right]_{4},\left[\mu_{\mp}\right]_{5}\right),
\end{align}
where the subscript $j$ of $\left[\mu_{\mp}\right]_{j}$, $j=1,\ldots,5$ stands for the $j$th column of $\mu_{\mp}$.

As $t\rightarrow\mp\infty$, it satisfies the following asyptotic conditions
\begin{align*}
\mu_{\mp}\rightarrow\mathbb{I}_{5},
\end{align*}
where $\mathbb{I}_{5}$ denotes a $5\times5$ identity matrix.

In fact, $J_{\mp}$ are solutions of the following Volterra integral equations
\begin{align}\label{af-6}
\mu_{\mp}(t,k)=\mathbb{I}_{5}+\int_{\mp\infty}^{t}\textrm{e}^{4\textrm{i}k^{3}
\widehat{\sigma}_{3}(t-\epsilon)}\widetilde{Q}(\epsilon)
\mu_{\mp}(\epsilon,k)\mathrm{d}\epsilon,
\end{align}
with $\textrm{e}^{\textrm{i}a
\widehat{\sigma}_{3}}\mathcal{A}=\textrm{e}^{\textrm{i}a
\sigma_{3}}\mathcal{A}\textrm{e}^{-\textrm{i}a
\sigma_{3}}$.
\vspace{0.2cm}

\noindent
\textbf{Assumption 2.1}
In order to ensure the sense of the equation \eqref{af-6}, we supplement
that
\begin{align}\nonumber
\int_{\mathbb{R}}\left(|q(x,t)|+|q_{x}(x,t)|+|q_{xx}(x,t)|\right)\mathrm{d}t<\infty,~~ \forall x\in\mathbb{R}.
\end{align}
Under the Assumption 2.1, we conclude the following proposition.
\vspace{0.2cm}

\noindent
\textbf{Proposition 2.2} For each $t\in \mathbb{R}$, the solutions $\left[\mu_{+}\right]_{1}$, $\left[\mu_{+}\right]_{2}$, $\left[\mu_{+}\right]_{3}$, $\left[\mu_{+}\right]_{4}$ and $\left[\mu_{-}\right]_{5}$ are analytically extended in $D_{+}$; the solutions $\left[\mu_{-}\right]_{1}$, $\left[\mu_{-}\right]_{2}$, $\left[\mu_{-}\right]_{3}$, $\left[\mu_{-}\right]_{4}$ and $\left[\mu_{+}\right]_{5}$ are analytically extended in $D_{-}$. Here $D_{+}=\left\{k\in\mathbb{C}|\mathrm{Re}k\mathrm{Im}k>0\right\}$ and
$D_{-}=\left\{k\in\mathbb{C}|\mathrm{Re}k\mathrm{Im}k<0\right\}$.
\vspace{0.2cm}

\noindent
\textbf{Proof.} Based on the Volterra integral equations \eqref{af-6}, it is easy to prove the existence, uniqueness and analyticity of the solutions $\mu_{\mp}$.
\vspace{0.2cm}

Let $\Sigma=\mathbb{R}\cup\textrm{i}\mathbb{R}$. Moreover, for $k\in\Sigma$, it follows from the trace $\mathrm{Tr}(\widetilde{Q})=0$ and the Volterra integral equations \eqref{af-6} that we know $\det \mu_{\mp}=1$.

There exists a scattering matrix $S(k)=(s_{ij}(k))_{5\times5}$ that is independent on $t$ and relates $\mu_{-}\textrm{e}^{4\textrm{i}k^{3}t}$ and $\mu_{+}\textrm{e}^{4\textrm{i}k^{3}t}$, i.e.,
\begin{align}\label{af-7}
\mu_{-}\textrm{e}^{4\textrm{i}k^{3}t}
=\mu_{+}\textrm{e}^{4\textrm{i}k^{3}t}S(k),~~k\in\Sigma,
\end{align}
and $\det S(k)=1$.
\vspace{0.2cm}

\noindent
\textbf{Proposition 2.3} The scattering coefficients $s_{1j}(k)$ and $s_{i1}(k)$, $i,j=1,2,3,4$ are analytically extended in $D_{-}$; the scattering coefficient $s_{55}(k)$ is analytically extended in $D_{+}$; the scattering coefficients $s_{5j}(k)$ and $s_{i5}(k)$, $i,j=1,2,3,4$ are generally defined on $\Sigma$.
\vspace{0.2cm}

\noindent
\textbf{Proof.} In terms of the expression \eqref{af-7}, it is obvious that properties of the scattering coefficients $s_{ij}(k)$, $j=1,\ldots,5$ and $k\in\Sigma$ can be deduced from the Proposition 2.2.
\vspace{0.2cm}

For $k\in D_{+}$, we define a $5\times 5$ matrix function. It is analytic and has the following form
\begin{align}\label{af-8}
\mathcal{P}_{1}:=\mathcal{P}_{1}(t,k)=\left(\left[\mu_{+}\right]_{1}, \left[\mu_{+}\right]_{2}, \left[\mu_{+}\right]_{3}, \left[\mu_{+}\right]_{4}, \left[\mu_{-}\right]_{5}\right).
\end{align}
Let $\mathcal{P}_{+}:=\mathcal{P}_{+}(t,k)$ represent the left limit of $\mathcal{P}_{1}$ in the set $\Sigma$. It follows from Eq. \eqref{af-7} that one arrives at
\begin{align}\label{af-9}
\mathcal{P}_{+}=\mu_{+}\left(
                         \begin{array}{ccccc}
                           1 & 0 & 0 & 0 & s_{15}(k)\textrm{e}^{8\textrm{i}k^{3}t} \\
                           0 & 1 & 0 & 0 & s_{25}(k)\textrm{e}^{8\textrm{i}k^{3}t} \\
                           0 & 0 & 1 & 0 & s_{35}(k)\textrm{e}^{8\textrm{i}k^{3}t} \\
                           0 & 0 & 0 & 1 & s_{45}(k)\textrm{e}^{8\textrm{i}k^{3}t} \\
                           0 & 0 & 0 & 0 & s_{55}(k) \\
                         \end{array}
                       \right).
\end{align}

Subsequently, we introduce the adjoint equation of the second expression in \eqref{af-3}, that is
\begin{align}\label{af-10}
\nu_{t}=4\textrm{i}k^{3}\left[\sigma_{3},\nu\right]+\nu\widetilde{Q}.
\end{align}
Then the matrix inverses of $\mu_{\mp}$ can be written as
\begin{align*}
\mu_{\mp}^{-1}=\left(\left[\mu_{\mp}^{-1}\right]^{1},\left[\mu_{\mp}^{-1}\right]^{2}
,\left[\mu_{\mp}^{-1}\right]^{3},\left[\mu_{\mp}^{-1}\right]^{4},
\left[\mu_{\mp}^{-1}\right]^{5}\right)^{\mathrm{T}},
\end{align*}
and it obeys Eq. \eqref{af-10}.

Similarly, we have the following proposition for $\mu_{\mp}^{-1}$.
\vspace{0.2cm}

\noindent
\textbf{Proposition 2.4} For each $t\in \mathbb{R}$, the solutions $\left[\mu_{+}^{-1}\right]^{1}$, $\left[\mu_{+}^{-1}\right]^{2}$, $\left[\mu_{+}^{-1}\right]^{3}$, $\left[\mu_{+}^{-1}\right]^{4}$ and $\left[\mu_{-}^{-1}\right]^{5}$ are analytically extended in $D_{-}$; the solutions $\left[\mu_{-}^{-1}\right]^{1}$, $\left[\mu_{-}^{-1}\right]^{2}$, $\left[\mu_{-}^{-1}\right]^{3}$, $\left[\mu_{-}^{-1}\right]^{4}$ and $\left[\mu_{+}^{-1}\right]^{5}$ are analytically extended in $D_{+}$.
\vspace{0.2cm}

According to the relations between
$\nu_{-}\textrm{e}^{4\textrm{i}k^{3}t}$ and $\nu_{+}\textrm{e}^{4\textrm{i}k^{3}t}$ as well as \eqref{af-7}, one has
\begin{align}\label{af-12}
\textrm{e}^{4\textrm{i}k^{3}t}\mu_{-}^{-1}
=R(k)\textrm{e}^{4\textrm{i}k^{3}t}\mu_{+}^{-1},~~k\in\Sigma.
\end{align}
Here $R(k)=S^{-1}(k)$ and $\det R(k)=1$.
\vspace{0.2cm}

\noindent
\textbf{Proposition 2.5} The scattering coefficients $r_{1j}(k)$ and $r_{i1}(k)$, $i,j=1,2,3,4$ are analytically extended in $D_{+}$; the scattering coefficient $r_{55}(k)$ is analytically extended in $D_{-}$; the scattering coefficients $r_{5j}(k)$ and $r_{i5}(k)$, $i,j=1,2,3,4$ are generally defined on $\Sigma$.
\vspace{0.2cm}

For $k\in D_{-}$, we define a new $5\times 5$ matrix function.

It is analytic and has the following form
\begin{align}\label{af-13}
\mathcal{P}_{2}:=\mathcal{P}_{2}(t,k)=\left(\left[\mu_{+}^{-1}\right]^{1}, \left[\mu_{+}^{-1}\right]^{2}, \left[\mu_{+}^{-1}\right]^{3}, \left[\mu_{+}^{-1}\right]^{4}, \left[\mu_{-}^{-1}\right]^{5}\right)^{\mathrm{T}}.
\end{align}
Let $\mathcal{P}_{-}:=\mathcal{P}_{-}(t,k)$ represent the right limit of $\mathcal{P}_{2}$ in the set $\Sigma$.

It follows from Eq. \eqref{af-13} that one concludes
\begin{align}\label{af-14}
\mathcal{P}_{-}=\left(
                         \begin{array}{ccccc}
                           1 & 0 & 0 & 0 & 0 \\
                           0 & 1 & 0 & 0 & 0 \\
                           0 & 0 & 1 & 0 & 0 \\
                           0 & 0 & 0 & 1 & 0 \\
                           r_{51}(k)\textrm{e}^{-8\textrm{i}k^{3}t} & r_{52}(k)\textrm{e}^{-8\textrm{i}k^{3}t} & r_{53}(k)\textrm{e}^{-8\textrm{i}k^{3}t} & r_{54}(k)\textrm{e}^{-8\textrm{i}k^{3}t} & r_{55}(k) \\
                         \end{array}
                       \right)\mu_{+}^{-1}.
\end{align}

\subsection{Asymptotic behaviors}

In this subsection, we shall study the asymptotic behaviors of $\mathcal{P}_{1}$ and $\mathcal{P}_{2}$. According to expressions \eqref{af-8} and \eqref{af-13}, we know that their asymptotic behaviors are defined by the asymptotic behaviors of the solutions $\mu_{\mp}$. The standard Wentzel-Kramers-Brillouin expansions \cite{Voros-1989} are used
to derive the asymptotic behaviors of the solutions $\mu_{\mp}$.

More specifically, we introduce
\begin{align}\label{af-15}
\mu_{\mp}(k)&=[\mu_{\mp}]^{0}+\frac{[\mu_{\mp}]^{1}}{k}+\frac{[\mu_{\mp}]^{2}}{k^{2}}
+\frac{[\mu_{\mp}]^{3}}{k^{3}}+O\left(\frac{1}{k^{3}}\right),~~k\rightarrow\infty,\notag\\
\mu_{\mp}^{-1}(k)&=[\mu_{\mp}^{-1}]_{0}+\frac{[\mu_{\mp}^{-1}]_{1}}{k}
+\frac{[\mu_{\mp}^{-1}]_{2}}{k^{2}}
+\frac{[\mu_{\mp}^{-1}]_{3}}{k^{3}}+O\left(\frac{1}{k^{3}}\right),~~k\rightarrow\infty.
\end{align}
By inserting the first equation in \eqref{af-15} into the second expression in \eqref{af-3}, one obtains these equations
\begin{align*}
4\textrm{i}\left[\sigma_{3},[\mu_{\mp}]^{0}\right]&=0,\notag\\
4\textrm{i}\left[\sigma_{3},[\mu_{\mp}]^{1}\right]&+4Q\mu_{0,\mp}=0,\notag\\
4\textrm{i}\left[\sigma_{3},[\mu_{\mp}]^{2}\right]&+4Q\mu_{1,\mp}
+2\textrm{i}\left(Q^{2}+Q_{x}\right)\sigma_{3}[\mu_{\mp}]^{0}=0,\notag\\
4\textrm{i}\left[\sigma_{3},[\mu_{\mp}]^{3}\right]&+4Q[\mu_{\mp}]^{2}
+2\textrm{i}\left(Q^{2}+Q_{x}\right)\sigma_{3}[\mu_{\mp}]^{1}
\notag\\
&+\left(Q_{x}Q-QQ_{x}-Q_{xx}+2Q^{3}\right)[\mu_{\mp}]^{0}=[\mu_{\mp,t}]^{0}.
\end{align*}
Successively, we deduce that $[\mu_{\mp,t}]^{0}=0$, so $[\mu_{\mp}]^{0}=\mathbb{I}_{5}$.

Similarly, it follows from the second equation in \eqref{af-15} that we calculate
\begin{align}\nonumber
[\mu_{\mp,t}^{-1}]_{0}=0,~~[\mu_{\mp}^{-1}]_{0}=\mathbb{I}_{5}.
\end{align}
Therefore the asymptotic behaviors of $\mathcal{P}_{1}$ and $\mathcal{P}_{2}$ are concluded as
\begin{align}\label{af-17}
&\mathcal{P}_{1}\rightarrow\mathbb{I}_{5},~~k\rightarrow\infty,\notag\\
&\mathcal{P}_{2}\rightarrow\mathbb{I}_{5},~~k\rightarrow\infty.
\end{align}

In order to get whole analyticity in $D_{+}$ and $D_{-}$, we have to construct a new matrix function $\mathcal{P}(t,k)$ for the ngSS system \eqref{ngss} in the light of two functions $\mathcal{P}_{+}(t,k)$ \eqref{af-9} and $\mathcal{P}_{-}(t,k)$ \eqref{af-14}. To clarify,
\begin{align*}
\mathcal{P}&(t,k)=\mathcal{P}_{-}(t,k)\mathcal{P}_{+}(t,k)\notag\\
&=\left(
                         \begin{array}{ccccc}
                           1 & 0 & 0 & 0 & s_{15}(k)\textrm{e}^{8\textrm{i}k^{3}t} \\
                           0 & 1 & 0 & 0 & s_{25}(k)\textrm{e}^{8\textrm{i}k^{3}t} \\
                           0 & 0 & 1 & 0 & s_{35}(k)\textrm{e}^{8\textrm{i}k^{3}t} \\
                           0 & 0 & 0 & 1 & s_{45}(k)\textrm{e}^{8\textrm{i}k^{3}t} \\
                           r_{51}(k)\textrm{e}^{-8\textrm{i}k^{3}t} & r_{52}(k)\textrm{e}^{-8\textrm{i}k^{3}t} & r_{53}(k)\textrm{e}^{-8\textrm{i}k^{3}t} & r_{54}(k)\textrm{e}^{-8\textrm{i}k^{3}t} & 1 \\
                         \end{array}
                       \right),~k\in\Sigma,
\end{align*}
and satisfies the asymptotic behaviors \eqref{af-17}.

\subsection{Spatial evolution}

According to the definitions of scattering matrices $S(k)$ in \eqref{af-7} and $R(k)$ in \eqref{af-12}, we know that
the continuous scattering data $s_{i5}$ and $r_{5j}$, $i,j=1,2,3,4$ are related to  parameters $x$. Afterwards we consider the spatial evolution.

Combining the first expression in \eqref{af-3} and these two scattering matrices can generate
\begin{align}\nonumber
S_{x}=\textrm{i}k\left[\sigma_{3},S\right],~~
R_{x}=\textrm{i}k\left[\sigma_{3},R\right].
\end{align}
More specifically,
\begin{equation*}
s_{i5,x}=2\textrm{i}ks_{i5},~~
r_{5j,x}=-2\textrm{i}kr_{5j},~j=1,2,3,4.
\end{equation*}

As a result, when $k\in\Sigma$, we have a new matrix function
\begin{small}
\begin{align}\label{af-19}
&\mathcal{P}(x,t,k)=\mathcal{P}_{-}(x,t,k)\mathcal{P}_{+}(x,t,k)\notag\\
&=\left(
                         \begin{array}{ccccc}
                           1 & 0 & 0 & 0 & s_{15}(0,k)\textrm{e}^{2\theta(k)} \\
                           0 & 1 & 0 & 0 & s_{25}(0,k)\textrm{e}^{2\theta(k)} \\
                           0 & 0 & 1 & 0 & s_{35}(0,k)\textrm{e}^{2\theta(k)} \\
                           0 & 0 & 0 & 1 & s_{45}(0,k)\textrm{e}^{2\theta(k)} \\
                           r_{51}(0,k)\textrm{e}^{-2\theta(k)} & r_{52}(0,k)\textrm{e}^{-2\theta(k)} & r_{53}(0,k)\textrm{e}^{-2\theta(k)} & r_{54}(0,k)\textrm{e}^{-2\theta(k)} & 1 \\
                         \end{array}
                       \right),
\end{align}
\end{small}
where $s_{i5}(0,k)$ and $r_{5j}(0,k)$ are the values of $s_{i5}$ and $r_{5j}$ at $x=0$, $i,j=1,2,3,4$,
\begin{align}\label{af-20}
\theta(k):=\theta(x,t,k)=\textrm{i}kx+4\textrm{i}k^{3}t.
\end{align}

\noindent
\textbf{Proposition 2.6}
The function $\mathcal{P}(x,t,k)$ satisfies the matrix RHP
\begin{itemize}
  \item Analyticity: $\mathcal{P}_{1}(x,t,k)$ and $\mathcal{P}_{2}(x,t,k)$ are analytic in $D_{+}$ and $D_{-}$, respectively.
  \item Canonical normalisation conditions:
  \begin{align*}
&\mathcal{P}_{1}(x,t,k)\rightarrow\mathbb{I}_{5},~~k\in D_{+}\rightarrow\infty,\notag\\
&\mathcal{P}_{2}(x,t,k)\rightarrow\mathbb{I}_{5},~~k\in
D_{-}\rightarrow\infty.
\end{align*}
\end{itemize}

\subsection{Symmetry reductions}

We here study the symmetry relations of the matrix functions $\mathcal{P}_{1}(x,t,k)$ and $\mathcal{P}_{2}(x,t,k)$. The symmetries of the two functions can be derived from ones of the solutions $\mu_{\mp}(x,t,k)$. In fact, the symmetry relations of $\mu_{\mp}(x,t,k)$ mainly rely on the potential matrix $Q$.

Now we explore the first symmetry reduction, that is
\begin{align}\label{af-22}
Q(x,t)=-\Lambda Q(-x,t)\Lambda,
\end{align}
with
\begin{align}\nonumber
\Lambda=\left(
          \begin{array}{ccccc}
            0 & 0 & 1 & 0 & 0 \\
            0 & 0 & 0 & 1 & 0 \\
            1 & 0 & 0 & 0 & 0 \\
            0 & 1 & 0 & 0 & 0 \\
            0 & 0 & 0 & 0 & -1 \\
          \end{array}
        \right).
\end{align}
On the basis of Eqs. \eqref{af-3} and \eqref{af-22}, we calculate the symmetry reduction of $\mu_{\mp}$,
\begin{align}\label{af-23}
\mu_{\mp}(x,t,k)=\Lambda \mu_{\mp}(-x,t,-k)\Lambda.
\end{align}
Uniting the matrix function $\mathcal{P}_{1}$ defined by \eqref{af-8} and Eq. \eqref{af-23}, the symmetry reduction of $\mathcal{P}_{1}$ is given by
\begin{align}\label{af-24}
\mathcal{P}_{1}(x,t,k)=\Lambda \mathcal{P}_{1}(-x,t,-k)\Lambda,~~k\in D_{+}.
\end{align}

Then we consider the second symmetry reduction
\begin{align}\label{af-25}
Q(x,t)=-Q^{\dagger}(x,t),
\end{align}
where the sign $``\dagger"$ stands for the Hermitian conjugation.

It follows from Eqs. \eqref{af-3} and \eqref{af-25} that the symmetry reduction of $\mu_{\mp}$ reads
\begin{align}\label{af-26}
\mu_{\mp}(x,t,k)=\left(\mu_{\mp}^{\dagger}(x,t,k^{\ast})\right)^{-1}.
\end{align}
On the consideration of Eqs. \eqref{af-13} and \eqref{af-26}, we write the symmetry reduction of $\mathcal{P}_{2}$ as
\begin{align}\label{af-27}
\mathcal{P}_{2}(x,t,k)=\mathcal{P}_{1}^{\dagger}(x,t,k^{\ast}),~~k\in D_{-}.
\end{align}

\section{Inverse scattering transform with simple pole}

In what follows, we need to derive the general zero structure of the RHP in Proposition 2.6.

\subsection{Explicit solution of the RHP}

According to the above symmetry reductions Eqs. \eqref{af-24} and \eqref{af-27}, we know that if $k$ is a simple zero of $\det\mathcal{P}_{1}$, then $-k$ is also a zero of $\det\mathcal{P}_{1}$. While $k^{\ast}$ and $-k^{\ast}$ both are zeros of $\det\mathcal{P}_{2}$.

Let $\widehat{k}$ denote a set containing $k^{\ast}$ and $-k^{\ast}$.

We suppose that $\det\mathcal{P}_{1}$ has a number
of $2 N$ simple zeros $\{k_{i}\}_{1}^{2N}$ in $D_{+}$, where $k_{N+i}=-k_{i}$ $(i= 1,2,\ldots,N)$. In a similar way, $\det\mathcal{P}_{2}$ also has
$2 N$ simple zeros $\left\{\widehat{k}_{j}\right\}_{1}^{2N}$ in $D_{-}$, and satisfy $\widehat{k}_{j}=k_{j}^{\ast}$, $(j = 1,2,\ldots,2N)$.

There are two vectors $U_{l} = \left(U_{l}(x,t)\right)_{5\times1}$ and $\widehat{U}_{l} =\left(\widehat{U}_{l}(x,t)\right)_{1\times 5}$, they respectively  span into $\ker[\mathcal{P}_{1}(k_{l})]$ and $\ker[\mathcal{P}_{2}(\widehat{k}_{l})]$, $l=,2,\ldots,2N$. Both $\ker[\mathcal{P}_{1}(k_{l})]$ and $\ker[\mathcal{P}_{2}(\widehat{k}_{l})]$ are 1D since the zeros of $\det\mathcal{P}_{1}$ and $\det\mathcal{P}_{2}$ always appear in quadruples.

As a result, we have
\begin{align}\label{af-28}
\mathcal{P}_{1}(k_{l})U_{l}=0,~~
\widehat{U}_{l}\mathcal{P}_{2}(\widehat{k}_{l})=0.
\end{align}
The discrete and continuous scattering data are severally given by
$\left\{k_{l},\widehat{k}_{l},U_{l},\widehat{U}_{l}\right\}$ and $\left\{s_{l5},s_{35},r_{51},r_{53}\right\}$.

First, we investigate the symmetry reductions of the discrete scattering data. In terms of definitions of simple zeros, we have
\begin{align*}
&k_{N+i} = -k_{i},~~1\leq i\leq N,\notag\\
&\widehat{k}_{j}=k_{j}^{\ast},~~1\leq j\leq 2N.
\end{align*}
Accordingly, the symmetry reductions of vectors $U_{i}$ and $\widehat{U}_{j}$ are
\begin{align}\label{af-30}
&U_{N+i}(x,t)= \Lambda U_{i}(-x,t),~~1\leq i\leq N,\notag\\
&\widehat{U}_{j}(x,t)=U_{j}^{\dagger}(x,t),~~1\leq j\leq 2N.
\end{align}
It follows from the first equation in \eqref{af-28} and Eq. \eqref{af-3} that we have
\begin{align}\label{af-31}
U_{l,x}(x,t)= \textrm{i}k_{l}U_{l},~~
U_{l,t}(x,t)= 4\textrm{i}k_{l}^{3}U_{l},~~1\leq l\leq N.
\end{align}
Therefore, by means of uniting expressions \eqref{af-30} and \eqref{af-31}, we deduce the final forms of these two vectors as follows
\begin{align}\label{af-32}
&U_{l}= \left\{\begin{array}{cc}
                 \textrm{e}^{\theta(x,t,k_{l})\sigma_{3}}U_{l,0}, & 1\leq l\leq N, \\
                 \Lambda\textrm{e}^{\theta(-x,t,k_{l-N})\sigma_{3}}U_{l-N,0}, & N+1\leq l\leq 2N,
               \end{array}\right.\notag\\
&\widehat{U}_{l}= \left\{\begin{array}{cc}
                 U_{l,0}^{\dagger}\textrm{e}^{\theta^{\ast}(x,t,k_{l})\sigma_{3}}, & 1\leq l\leq N, \\
                 U_{l-N,0}^{\dagger}\textrm{e}^{\theta^{\ast}(-x,t,k_{l-N})\sigma_{3}}\Lambda, & N+1\leq l\leq 2N,
               \end{array}\right.
\end{align}
where $U_{l,0}\in \mathbb{C}$, and $\theta(x,t,k_{l})$ is defined by Eq. \eqref{af-20} through replacing $k$ with $k_{l}$, ($k_{l}\in D_{+}$).

In order to regular the RHP in Proposition 2.6 with the specific zero
structure to a regular one without zeros, two dressing matrices are presented as follows
\begin{align*}
&\Gamma_{1}(k)=\mathbb{I}_{5}-\sum_{i=1}^{2N}\sum_{l=1}^{2N}
\frac{U_{i}\widehat{U}_{l}\left(M^{-1}\right)_{il}}
{k-\widehat{k}_{l}},~~k\in D_{+}\cup\Sigma,\notag\\
&\Gamma_{2}(k)=\mathbb{I}_{5}+\sum_{i=1}^{2N}\sum_{l=1}^{2N}
\frac{U_{i}\widehat{U}_{l}\left(M^{-1}\right)_{il}}
{k-k_{i}},~~k\in D_{-}\cup\Sigma.
\end{align*}
Here $M=(m_{il})_{2N\times2N}$, and
\begin{align}\label{af-34}
m_{il}=\frac{\widehat{U}_{i}U_{l}}{k_{l}-\widehat{k}_{i}},~~1\leq i,l\leq2N.
\end{align}
Furthermore, the RHP \eqref{af-19} is formally solved as
\begin{align*}
&\mathcal{P}_{1}(k)=\widehat{\mathcal{P}}_{1}(k)\Gamma_{1}(k),\notag\\
&\mathcal{P}_{2}(k)=\Gamma_{2}(k)\widehat{\mathcal{P}}_{2}(k),
\end{align*}
where $\widehat{\mathcal{P}}_{1}(k)$ and $\widehat{\mathcal{P}}_{2}(k)$ can be characterized by the following integral equations
\begin{align*}
\left(\widehat{\mathcal{P}}_{1}(k)\right)^{-1}
&=\mathbb{I}_{5}+\frac{1}{2\pi\textrm{i}}\int_{\Sigma}
\frac{\Gamma_{1}(\zeta)J(\zeta)\Gamma_{1}(\zeta)
\left(\widehat{\mathcal{P}}^{+}(\zeta)\right)^{-1}}
{\zeta-k}\mathrm{d}\zeta,~~k\in D_{+},\notag\\
\widehat{\mathcal{P}}_{2}(k)
&=\mathbb{I}_{5}+\frac{1}{2\pi\textrm{i}}\int_{\Sigma}
\frac{\Gamma_{1}(\zeta)J(\zeta)\Gamma_{2}(\zeta)
\widehat{\mathcal{P}}^{-}(\zeta)}
{\zeta-k}\mathrm{d}\zeta,~~k\in D_{-},
\end{align*}
where $\widehat{\mathcal{P}}^{+}$ and $\widehat{\mathcal{P}}^{-}$ are left and right  limits of $\widehat{\mathcal{P}}_{1}$ and $\widehat{\mathcal{P}}_{2}$ in the set $\Sigma$, respectively.

The matrix function
\begin{small}
\begin{align*}
&J(\zeta)=\notag\\
&\left(
                         \begin{array}{ccccc}
                           0 & 0 & 0 & 0 & -s_{15}(0,\zeta)\textrm{e}^{2\theta(\zeta)} \\
                           0 & 0 & 0 & 0 & -s_{25}(0,\zeta)\textrm{e}^{2\theta(\zeta)} \\
                           0 & 0 & 0 & 0 & -s_{35}(0,\zeta)\textrm{e}^{2\theta(\zeta)} \\
                           0 & 0 & 0 & 0 & -s_{45}(0,\zeta)\textrm{e}^{2\theta(\zeta)} \\
                           -r_{51}(0,\zeta)\textrm{e}^{-2\theta(\zeta)} & -r_{52}(0,\zeta)\textrm{e}^{-2\theta(\zeta)} & -r_{53}(0,\zeta)\textrm{e}^{-2\theta(\zeta)} & -r_{54}(0,\zeta)\textrm{e}^{-2\theta(\zeta)} & 0 \\
                         \end{array}
                       \right).
\end{align*}
\end{small}
We next need to consider the reflectionless condition. That is to say, $s_{i5}(0,k)$ and $r_{5j}(0,k)$, $i,j=1,2,3,4$ all equal to 0, so $J(k)$ turns into the zero matrix. Then $\widehat{\mathcal{P}}_{1}(k)$ and $\widehat{\mathcal{P}}_{2}(k)$ reduce to two $5\times5$ identity matrices.

Furthermore, we calculate the explicit solutions as
\begin{align}\label{af-38}
&\mathcal{P}_{1}(k)=\mathbb{I}_{5}-\sum_{i=1}^{2N}\sum_{l=1}^{2N}
\frac{U_{i}\widehat{U}_{l}\left(M^{-1}\right)_{il}}
{k-\widehat{k}_{l}},~~k\in D_{+}\cup\Sigma,\notag\\
&\mathcal{P}_{2}(k)=\mathbb{I}_{5}+\sum_{i=1}^{2N}\sum_{l=1}^{2N}
\frac{U_{i}\widehat{U}_{l}\left(M^{-1}\right)_{il}}
{k-k_{i}},~~k\in D_{-}\cup\Sigma.
\end{align}

\subsection{$N$-soliton solution}

With the help of the standard Wentzel-Kramers-Brillouin expansions \cite{Voros-1989}, we have
\begin{align}\label{af-39}
\mathcal{P}_{1}(k)&=\mathbb{I}_{5}+\frac{\mathcal{P}_{1}^{[1]}}{k}
+O\left(\frac{1}{k}\right),~~k\in D_{+}\cup\Sigma\rightarrow\infty, \notag\\
\mathcal{P}_{2}(k)&=\mathbb{I}_{5}+\frac{\mathcal{P}_{2}^{[1]}}{k}
+O\left(\frac{1}{k}\right),~~k\in D_{-}\cup\Sigma\rightarrow\infty.
\end{align}
By substituting the formulae in \eqref{af-39} into the first equation in \eqref{af-38}, one calculates
\begin{align}\label{af-40}
\mathcal{P}_{1}^{[1]}&=-\sum_{i=1}^{2N}\sum_{l=1}^{2N}
\frac{U_{i}\widehat{U}_{l}\left(M^{-1}\right)_{il}}
{k-\widehat{k}_{l}},\notag\\
\mathcal{P}_{2}^{[1]}&=\sum_{i=1}^{2N}\sum_{l=1}^{2N}
\frac{U_{i}\widehat{U}_{l}\left(M^{-1}\right)_{il}}
{k-k_{i}}.
\end{align}
Combining \eqref{af-39} and the second equation in \eqref{af-3} yields that
\begin{align}\nonumber
Q=-\textrm{i}\left[\sigma_{3},\mathcal{P}_{1}^{[1]}\right]
=-\textrm{i}\left[\sigma_{3},\mathcal{P}_{2}^{[1]}\right].
\end{align}
Consequently, the general solutions are given by
\begin{align}\label{af-41}
q(x,t)&=-2\textrm{i}\left(\mathcal{P}_{1}^{[1]}\right)_{15}
=2\textrm{i}\left(\mathcal{P}_{2}^{[1]}\right)_{15}
,~~
-q^{\ast}(x,t)
=2\textrm{i}\left(\mathcal{P}_{1}^{[1]}\right)_{51}
=-2\textrm{i}\left(\mathcal{P}_{2}^{[1]}\right)_{51},\notag\\
q^{\ast}(x,t)&=-2\textrm{i}\left(\mathcal{P}_{1}^{[1]}\right)_{25}
=2\textrm{i}\left(\mathcal{P}_{2}^{[1]}\right)_{25}
,~~
-q(x,t)
=2\textrm{i}\left(\mathcal{P}_{1}^{[1]}\right)_{52}
=-2\textrm{i}\left(\mathcal{P}_{2}^{[1]}\right)_{52},\notag\\
q(-x,t)&=-2\textrm{i}\left(\mathcal{P}_{1}^{[1]}\right)_{35}
=2\textrm{i}\left(\mathcal{P}_{2}^{[1]}\right)_{35}
,~~
-q^{\ast}(-x,t)
=2\textrm{i}\left(\mathcal{P}_{1}^{[1]}\right)_{53}
=-2\textrm{i}\left(\mathcal{P}_{2}^{[1]}\right)_{53},\notag\\
q^{\ast}(-x,t)&=-2\textrm{i}\left(\mathcal{P}_{1}^{[1]}\right)_{45}
=-2\textrm{i}\left(\mathcal{P}_{2}^{[1]}\right)_{45}
,~~
-q(-x,t)
=2\textrm{i}\left(\mathcal{P}_{1}^{[1]}\right)_{54}
=-2\textrm{i}\left(\mathcal{P}_{2}^{[1]}\right)_{54},
\end{align}
where $\left(\mathcal{P}_{l}^{[1]}\right)_{ij}$ means the $i$th row and the $j$th column of $\mathcal{P}_{l}^{[1]}$, $l=1,2$.
\vspace{0.2cm}

\noindent
\textbf{Remark 3.1} In terms of two symmetries \eqref{af-24} and \eqref{af-27}, it is easy to verify the consistency of these general solutions given by \eqref{af-41} via some algebraic calculations.
\vspace{0.2cm}

Let the initial vector denote
\begin{align}\nonumber
U_{l,0}=\left(a_{l},b_{l},c_{l},d_{l},1\right)^{\mathrm{T}}\in\mathbb{C},~~1\leq l\leq N.
\end{align}
In order to obtain the general $N$-soliton solution, we find that the initial vector contains $5N$ free parameters, they are
$a_{1}, a_{2}, \ldots, a_{N}$, $b_{1}, b_{2}, \ldots, b_{N}$, $c_{1}, c_{2}, \ldots, c_{N}$, $d_{1}, d_{2}, \ldots, d_{N}$, $k_{1}, k_{2}, \ldots, k_{N}$.

According to the definition in Eq. \eqref{af-34}, we compute that

\noindent
when $1\leq i,l\leq N$,
\begin{align}\label{af-42}
m_{il}=\frac{\left(a_{i}^{\ast}a_{l}+b_{i}^{\ast}b_{l}
                +c_{i}^{\ast}c_{l}+d_{i}^{\ast}d_{l}\right)
                \textrm{e}^{\theta^{\ast}(k_{i})+\theta(k_{l})}+
                \textrm{e}^{-\left(\theta^{\ast}(k_{i})+\theta(k_{l})\right)}}
                {k_{l}-k_{i}^{\ast}},
\end{align}
when $1\leq i\leq N$ and $N+1\leq l\leq 2N$,
\begin{align}\label{af-43}
m_{il}=\frac{\left(a_{i}^{\ast}c_{l-N}+b_{i}^{\ast}d_{l-N}
                +c_{i}^{\ast}a_{l-N}+d_{i}^{\ast}b_{l-N}\right)
                \textrm{e}^{\theta^{\ast}(k_{i})+\theta(-x,t,k_{l-N})}-
                \textrm{e}^{-\left(\theta^{\ast}(k_{i})+\theta(-x,t,k_{l-N})\right)}}
                {-k_{l-N}-k_{i}^{\ast}},
\end{align}
when $N+1\leq i\leq 2N$ and $1\leq l\leq N$,
\begin{align}\label{af-44}
m_{il}=\frac{\left(c_{i-N}^{\ast}a_{l}+d_{i-N}^{\ast}b_{l}
                +a_{i-N}^{\ast}c_{l}+b_{i-N}^{\ast}d_{l}\right)
                \textrm{e}^{\theta^{\ast}(-x,t,k_{i-N})+\theta(k_{l})}-
                \textrm{e}^{-\left(\theta^{\ast}(-x,t,k_{i-N})+\theta(k_{l})\right)}}
                {k_{l}+k_{i-N}^{\ast}},
\end{align}
when $N+1\leq i,l\leq 2N$,
\begin{align}\label{af-45}
m_{il}=&\frac{\left(c_{i-N}^{\ast}c_{l-N}+d_{i-N}^{\ast}d_{l-N}
                +a_{i-N}^{\ast}a_{l-N}+b_{i-N}^{\ast}b_{l-N}\right)
                \textrm{e}^{\theta^{\ast}(-x,t,k_{i-N})+\theta(-x,t,k_{l-N})}}
                {-k_{l-N}+k_{i-N}^{\ast}}\notag\\
                &+\frac{\textrm{e}^{-\left(\theta^{\ast}(-x,t,k_{i-N})
                +\theta(-x,t,k_{l-N})\right)}}
                {-k_{l-N}+k_{i-N}^{\ast}}.
\end{align}
Uniting Eqs. \eqref{af-32}, \eqref{af-40} and \eqref{af-41} yields the following theorem.
\vspace{0.2cm}

\noindent
\textbf{Theorem 3.2}
The general $N$-soliton solution of the ngSS equation \eqref{ngss} with compact form is written as
\begin{align}\label{af-46}
q(x,t)=-2\textrm{i}\frac{\det H}{\det M}.
\end{align}
Here $M=\left(m_{il}\right)_{2N\times2N}$, and $m_{il}$ are given by Eqs. \eqref{af-42}-\eqref{af-45}. The matrix function $H$ is made up of two $2N\times 1$ matrices $\chi$ and $\omega$ as well as the matrix $M$,
\begin{align}\label{af-47}
H&=\left(
    \begin{array}{cc}
      0 & \chi^{\mathrm{T}} \\
      \omega & M \\
    \end{array}
  \right),\notag\\
\chi&=\left(a_{1}\textrm{e}^{\theta(k_{1})},a_{2}\textrm{e}^{\theta(k_{2})},\ldots,
a_{N}\textrm{e}^{\theta(k_{N})},c_{1}\textrm{e}^{\theta(-x,t,k_{1})},
c_{2}\textrm{e}^{\theta(-x,t,k_{2})},\ldots,
c_{N}\textrm{e}^{\theta(-x,t,k_{N})}\right)^{\mathrm{T}},\notag\\
\omega&=\left(\textrm{e}^{-\theta^{\ast}(k_{1})},\textrm{e}^{-\theta^{\ast}(k_{2})},\ldots,
\textrm{e}^{-\theta^{\ast}(k_{N})},-\textrm{e}^{-\theta^{\ast}(-x,t,k_{1})},
-\textrm{e}^{-\theta^{\ast}(-x,t,k_{2})},\ldots,
-\textrm{e}^{-\theta^{\ast}(-x,t,k_{N})}\right)^{\mathrm{T}}.
\end{align}

\subsection{$1$-soliton dynamics}

We let $N=1$, $b_{l}=a_{l}^{\ast}$ and $d_{l}=c_{l}^{\ast}$. It is noted that all derivations in this subsection are performed under this condition.

According to the expression \eqref{af-46}, the $1$-soliton solution is written as
\begin{align}\label{af-48}
q(x,t)=&2\textrm{i}\frac{a_{1}\left(m_{22}\textrm{e}^{\theta(k_{1})
-\theta^{\ast}(k_{1})}+m_{12}\textrm{e}^{\theta(k_{1})
-\theta^{\ast}(-x,t,k_{1})}\right)}{m_{11}m_{22}-m_{12}m_{21}}\notag\\
&-\frac{c_{1}\left(m_{21}\textrm{e}^{\theta(-x,t,k_{1})
-\theta^{\ast}(k_{1})}+m_{11}\textrm{e}^{\theta(-x,t,k_{1})
-\theta^{\ast}(-x,t,k_{1})}\right)}{m_{11}m_{22}-m_{12}m_{21}},
\end{align}
with
\begin{align}\label{af-49}
m_{11}=&\frac{2(|a_{1}|^{2}+|c_{1}|^{2})\textrm{e}^{\theta^{\ast}(k_{1})
+\theta(k_{1})}+\textrm{e}^{-\left(\theta^{\ast}(k_{1})+\theta(k_{1})\right)}}
{k_{1}-k_{1}^{\ast}},\notag\\
m_{12}=&\frac{2(a_{1}^{\ast}c_{1}+c_{1}^{\ast}a_{1})\textrm{e}^{\theta^{\ast}(k_{1})
+\theta(-x,t,k_{1})}-\textrm{e}^{-\left(\theta^{\ast}(k_{1})+\theta(-x,t,k_{1})\right)}}
{-k_{1}-k_{1}^{\ast}},\notag\\
m_{21}=&\frac{2(a_{1}^{\ast}c_{1}+c_{1}^{\ast}a_{1})\textrm{e}^{\theta^{\ast}(-x,t,k_{1})
+\theta(k_{1})}-\textrm{e}^{-\left(\theta^{\ast}(-x,t,k_{1})+\theta(k_{1})\right)}}
{k_{1}+k_{1}^{\ast}},\notag\\
m_{22}=&\frac{2(|a_{1}|^{2}+|c_{1}|^{2})\textrm{e}^{\theta^{\ast}(-x,t,k_{1})
+\theta(-x,t,k_{1})}+\textrm{e}^{-\left(\theta^{\ast}(-x,t,k_{1})+\theta(-x,t,k_{1})\right)}}
{-k_{1}+k_{1}^{\ast}}.
\end{align}
To simplify the following expressions, we let
\begin{align}\label{af-52}
\Delta_{1}=|a_{1}|^{2}+|c_{1}|^{2},~~
\Delta_{2}=a_{1}^{\ast}c_{1}+c_{1}^{\ast}a_{1}.
\end{align}
It follows from \eqref{af-49} that we calculate
\begin{align}\label{af-50}
m_{11}m_{22}-m_{12}m_{21}=&\frac{\left[2\Delta_{1}\textrm{e}^{\theta^{\ast}(k_{1})
+\theta(k_{1})}+\textrm{e}^{-\left(\theta^{\ast}(k_{1})+\theta(k_{1})\right)}\right]}
{4\Im^{2}k_{1}}\notag\\
&
\cdot
\frac{\left[2\Delta_{1}\textrm{e}^{\theta^{\ast}(-x,t,k_{1})
+\theta(-x,t,k_{1})}+\textrm{e}^{-\left(\theta^{\ast}(-x,t,k_{1})
+\theta(-x,t,k_{1})\right)}\right]}
{4\Im^{2}k_{1}}\notag\\
&+\frac{\left|2\Delta_{2}\textrm{e}^{\theta^{\ast}(k_{1})
+\theta(-x,t,k_{1})}-\textrm{e}^{-\left(\theta^{\ast}(k_{1})
+\theta(-x,t,k_{1})\right)}\right|^{2}}
{4\Re^{2}k_{1}}.
\end{align}
From Eq. \eqref{af-50}, the parameters $a_{1}$ and $c_{1}$ are not all 0. For this reason, there are three cases of values of $a_{1}$ and $c_{1}$:
\begin{description}
  \item[(1)] $a_{1}=0$ and $c_{1}\neq0$;
  \item[(2)] $a_{1}\neq0$ and $c_{1}=0$;
  \item[(3)] $a_{1}\neq0$ and $c_{1}\neq0$.
\end{description}

Under the condition of $\Re(\theta_{1}) = O(1)$, by directly computing, we further obtain the following forms
\begin{align}\label{af-51}
&q(x,t)\rightarrow\textrm{i}\left(k_{1}-k_{1}^{\ast}\right)\gamma_{\mp}
\textrm{e}^{\theta(k_{1})-\theta^{\ast}(k_{1})}
\mathrm{sech}\left(\theta(k_{1})+\theta^{\ast}(k_{1})+\delta_{\mp}\right),
~~t\rightarrow\mp\infty,
\end{align}
with
\begin{equation}\nonumber
\begin{aligned}
\gamma_{-}&=\frac{|k_{1}|(k_{1}+k_{1}^{\ast})a_{1}}{k_{1}
|k_{1}+k_{1}^{\ast}|\Delta_{1}},\notag\\
\gamma_{+}&=\frac{a_{1}+(k_{1}-k_{1}^{\ast})(k_{1}+k_{1}^{\ast})^{-1}
\Delta_{2}\Delta_{1}^{-1}c_{1}}{\sqrt{\Delta_{1}
(k_{1}-k_{1}^{\ast})^{2}(k_{1}+k_{1}^{\ast})^{-2}
\Delta_{2}^{2}\Delta_{1}^{-1}}},\notag\\
\delta_{-}&=\frac{1}{2}\ln\left[\frac{(k_{1}+k_{1}^{\ast})^{2}}
{4|k_{1}|^{2}}\cdot\Delta_{1}\right],\notag\\
\delta_{+}&=\frac{1}{2}\ln\left[\Delta_{1}-\frac{(k_{1}-k_{1}^{\ast})^{2}}
{(k_{1}+k_{1}^{\ast})^{2}}\cdot\frac{\Delta_{2}^{2}}{\Delta_{1}}\right].
\end{aligned}
\end{equation}
In consideration of $k_{1}\in\mathbb{C}$, we let $k_{1}=k_{1R}+\textrm{i}k_{1I}$, and $k_{1R}k_{1I}>0$.
Then Eq. \eqref{af-51} is reformulated as
\begin{align}\label{af-53}
|q(x,t)|\rightarrow&\frac{2|k_{1I}|\cdot|a_{1}|}{\sqrt{\Delta_{1}}}\cdot
\mathrm{sech}\left\{-2k_{1I}\left[x+4\left(3k_{1R}^{2}-k_{1I}^{2}\right)t\right]
+\delta_{-}\right\},~~t\rightarrow-\infty,\notag\\
|q(x,t)|\rightarrow&\frac{2|k_{1I}|\cdot|\Delta_{1}a_{1}k_{1R}+
\textrm{i}\Delta_{2}c_{1}k_{1I}|}{\sqrt{\Delta_{1}^{2}k_{1R}^{2}+
\Delta_{2}^{2}k_{1I}^{2}}\cdot\sqrt{\Delta_{1}}}\cdot
\mathrm{sech}\left\{-2k_{1I}\right.\notag\\
&\left.\left[x+4\left(3k_{1R}^{2}-k_{1I}^{2}\right)t\right]
+\delta_{+}\right\},~~t\rightarrow+\infty.
\end{align}

For the purpose of describing $1$-soliton dynamics more intuitively, we analyze the asymptotic behaviors of the $1$-soliton solution \eqref{af-48} in the above three cases in theory and draw corresponding figures from aspects of 3D profiles, projection profiles and wave propagations along the $x$ axis.

For the case (1), the asymptotic expressions of $q(x,t)$ in \eqref{af-53} are rewritten as
\begin{align}\nonumber
|q(x,t)|\rightarrow0,~~t\rightarrow\mp\infty.
\end{align}
This implies that the two asymptotic soliton degenerates to zero as $t$ goes to infinity. There is no asymptotic soliton in the moving frame with velocity $4\left(3k_{1R}^{2}-k_{1I}^{2}\right)$ as $t\rightarrow\mp\infty$.

Then we demonstrate the dynamic behavior for $1$-soliton solution by choosing $a_{1}=0$, $c_{1}=1$ and $k_{1}=0.01+0.5\textrm{i}$.
\vspace{0.2cm}

{\rotatebox{0}{\includegraphics[width=6.2cm,height=4.2cm,angle=0]{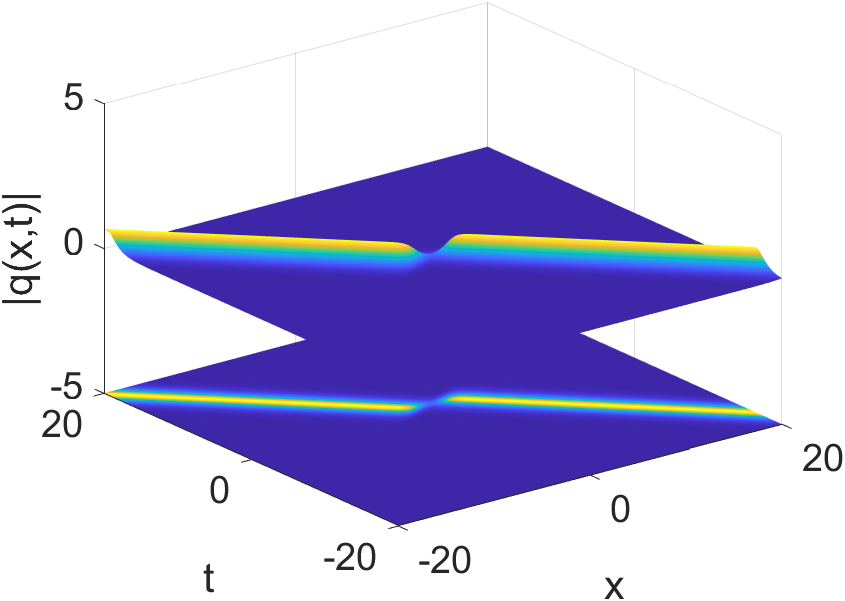}}}
~~{\rotatebox{0}{\includegraphics[width=5.4cm,height=4.2cm,angle=0]{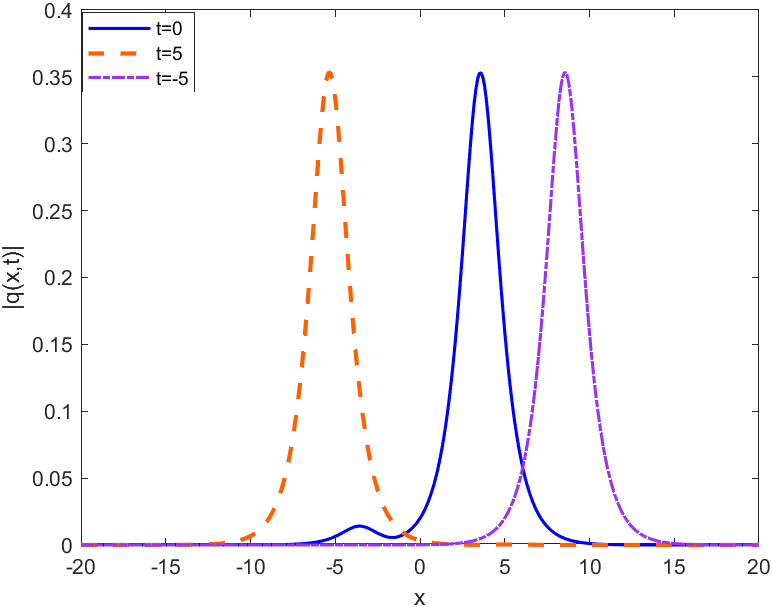}}}
\\
$~~~~~~~~~~~~~~~~~~~~~~~~~~~\textbf{(a)}
~~~~~~~~~~~~~~~~~~~~~~~~~~~~~~~~~~~~~~~~~~~~~~~\textbf{(b)}
$
\\

{\small \textbf{Figure 1.} The 1-soliton solution $q(x,t)$ in \eqref{af-53}:
(a) 3D and projection profiles;
(b) wave propagation along the $x$ axis at $t=0$ and $t=\pm5$.}
\vspace{0.2cm}

For the case (2), $\Delta_{1}=2|a_{1}|^{2}$ and $\Delta_{2}=0$. So the asymptotic expressions of $q(x,t)$ in \eqref{af-53} change to
\begin{equation}\nonumber
\begin{aligned}
|q(x,t)|\rightarrow\sqrt{2}|k_{1I}|\cdot
\mathrm{sech}\left\{-2k_{1I}\left[x+4\left(3k_{1R}^{2}-k_{1I}^{2}\right)t\right]
+\delta_{\mp}\right\},~~t\rightarrow\mp\infty,
\end{aligned}
\end{equation}
with
\begin{equation}\nonumber
\begin{aligned}
\delta_{-}=\ln\left(\frac{\sqrt{2}|a_{1}k_{1R}|}{\sqrt{k_{1R}^{2}+k_{1I}^{2}}}\right),
~~\delta_{+}=\ln\left(\sqrt{2}|a_{1}|\right).
\end{aligned}
\end{equation}
We note that the position shift is
\begin{equation}\nonumber
\begin{aligned}
\delta_{+}-\delta_{-}=\frac{1}{2}\ln\left(1+\frac{k_{1I}^{2}}{k_{1R}^{2}}\right),
\end{aligned}
\end{equation}
and the asymptotic soliton keeps its amplitude invariant except the position shift.

In what follows, we depict the dynamical pattern of $1$-soliton solution with $a_{1}=1$, $c_{1}=0$ and $k_{1}=1.6+\textrm{i}$.
\vspace{0.2cm}

{\rotatebox{0}{\includegraphics[width=6.2cm,height=4.2cm,angle=0]{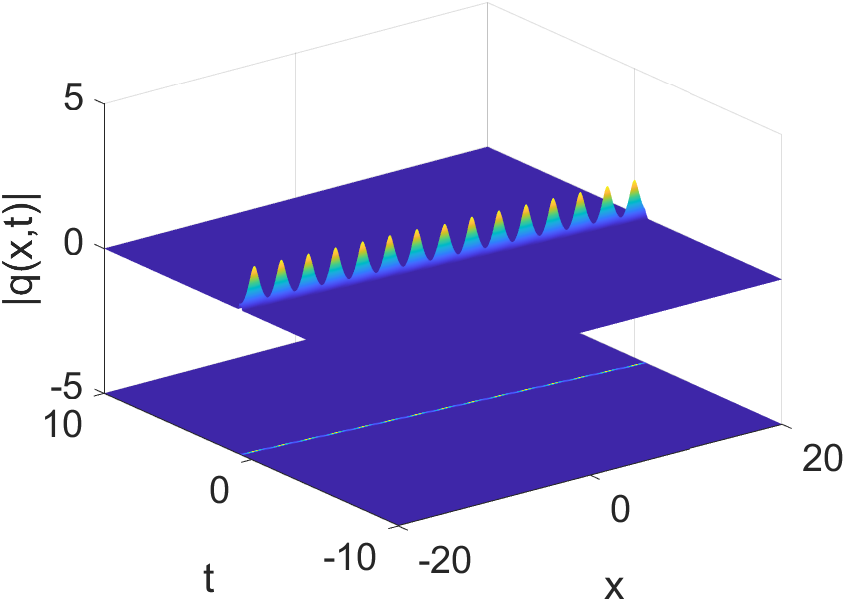}}}
~~{\rotatebox{0}{\includegraphics[width=5.4cm,height=4.2cm,angle=0]{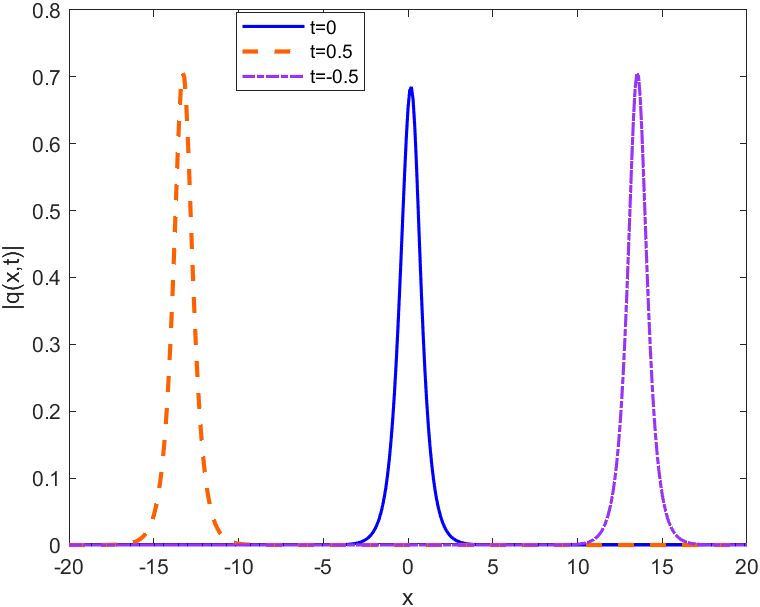}}}
\\
$~~~~~~~~~~~~~~~~~~~~~~~~~~~\textbf{(a)}
~~~~~~~~~~~~~~~~~~~~~~~~~~~~~~~~~~~~~~~~~~~~~~~\textbf{(b)}
$
\\

{\small \textbf{Figure 2.} The 1-soliton solution $q(x,t)$ in \eqref{af-53}:
(a) 3D and projection profiles;
(b) wave propagation along the $x$ axis at $t=0$ and $t=\pm0.5$.}
\vspace{0.2cm}

For the case (3), it follows from \eqref{af-53} that $|q(x,t)|$ goes to a single-soliton solution of hyperbolic secant type in the coordinate frame $\left(x+4\left(3k_{1R}^{2}-k_{1I}^{2}\right)t,t\right)$. The position shift is related to values of $k_{1R}$, $k_{1I}$, $a_{1}$, and $c_{1}$.
\vspace{0.2cm}

\noindent
\textbf{Proposition 3.3} Under the condition of $b_{1}=a_{1}^{\ast}$ and $d_{1}=c_{1}^{\ast}$. The amplitudes before and after the collision are severally denoted by $B$ and $A$, their expressions are as follows
\begin{equation}\nonumber
\begin{aligned}
&B:=\frac{2|k_{1I}|\cdot|a_{1}|}{\sqrt{\Delta_{1}}},\notag\\
&A:=\frac{2|k_{1I}|\cdot|\Delta_{1}a_{1}k_{1R}+
\textrm{i}\Delta_{2}c_{1}k_{1I}|}{\sqrt{\Delta_{1}^{2}k_{1R}^{2}+
\Delta_{2}^{2}k_{1I}^{2}}\cdot\sqrt{\Delta_{1}}}.
\end{aligned}
\end{equation}
Moreover, $A=B$ if and only if $\Re\left(a_{1}^{\ast}c_{1}\right)=0$ or $\Delta_{2}\left(|a_{1}|^{2}-|c_{1}|^{2}\right)k_{1I}=\textrm{i}
\left(a_{1}^{\ast}c_{1}-a_{1}c_{1}^{\ast}\right)\Delta_{1}k_{1R}$.
Here $\Delta_{1}$ and $\Delta_{2}$ are defined by \eqref{af-52}.
\vspace{0.2cm}

As we all know, for the generalized SS equation \cite{Geng-2016,Geng-2017}, its position shift is only dependent of the value of $k_{1}$, and the soliton are invariant before and after collision.
These characters are different from ones of the ngSS equation \eqref{ngss} that we have investigated above.

Based on the discussions above in case (3), we show some remarkable $1$-soliton solution dynamics. Here Figs. 3 and 4 are drawn on the condition of $a_{1}=\frac{\textrm{i}}{2}$, $c_{1}=\frac{\sqrt{3}\textrm{i}}{2}$. In Fig. 3, there is the interaction between two breathers. Fig. 4 displays the interaction between two $1$-solitons. In both figures the solitons change after the collision. Different from Figs. 3 and 4, two $1$-solitons keep travelling in their original directions and amplitudes after the collision in Fig. 5. That is to say, when $|a_{1}|=|c_{1}|$, they maintain the original features propagation after the collision by observing the pattern of the wave propagation along the $x$ axis at $t=\pm5$ in Fig. 5(b). Thus we learn that the values of parameters $a_{1}$ and $c_{1}$ determine whether the soliton changes after the collision.
\vspace{0.2cm}

{\rotatebox{0}{\includegraphics[width=6.2cm,height=4.2cm,angle=0]{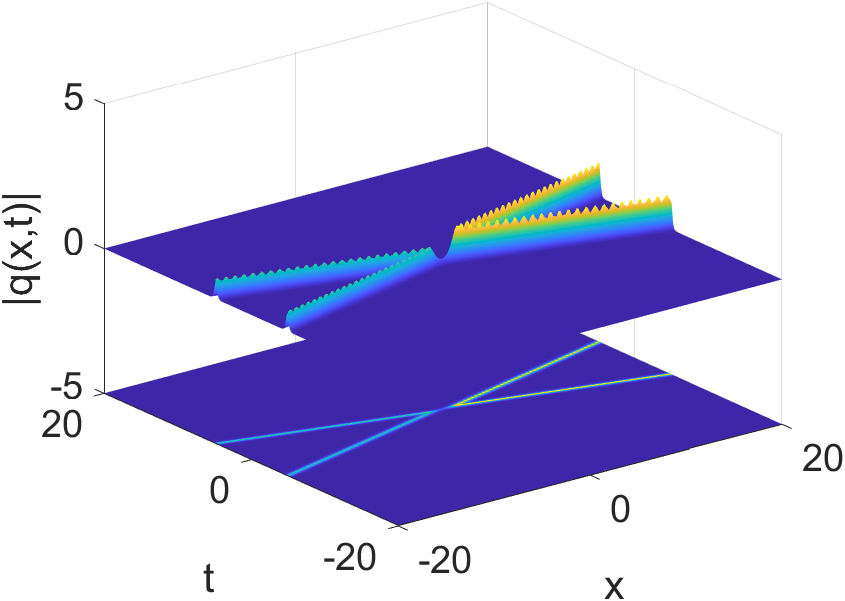}}}
~~{\rotatebox{0}{\includegraphics[width=5.4cm,height=4.2cm,angle=0]{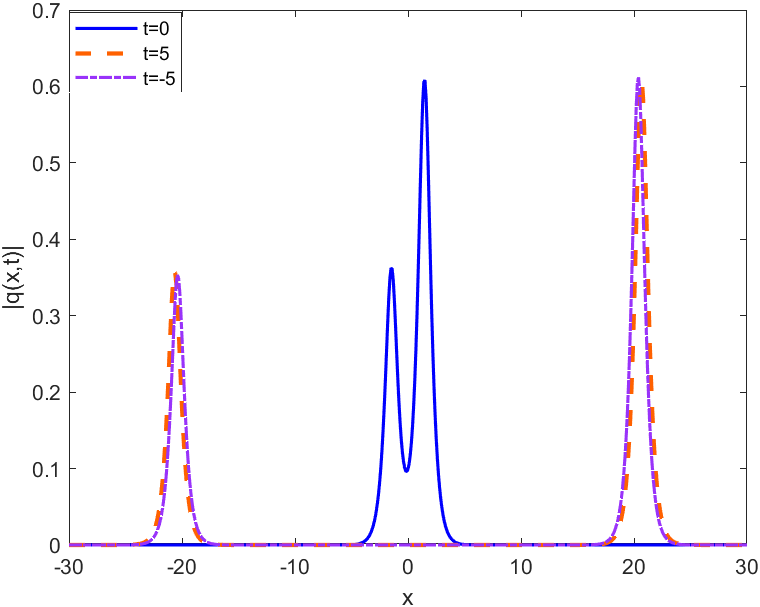}}}
\\
$~~~~~~~~~~~~~~~~~~~~~~~~~~~\textbf{(a)}
~~~~~~~~~~~~~~~~~~~~~~~~~~~~~~~~~~~~~~~~~~~~~~~\textbf{(b)}
$
\\

{\small \textbf{Figure 3.} The 1-soliton solution $q(x,t)$ in \eqref{af-53} with $k_{1}=0.1+\textrm{i}$:
(a) 3D and projection profiles;
(b) wave propagation along the $x$ axis at $t=0$ and $t=\pm5$.}
\vspace{0.2cm}

{\rotatebox{0}{\includegraphics[width=6.2cm,height=4.2cm,angle=0]{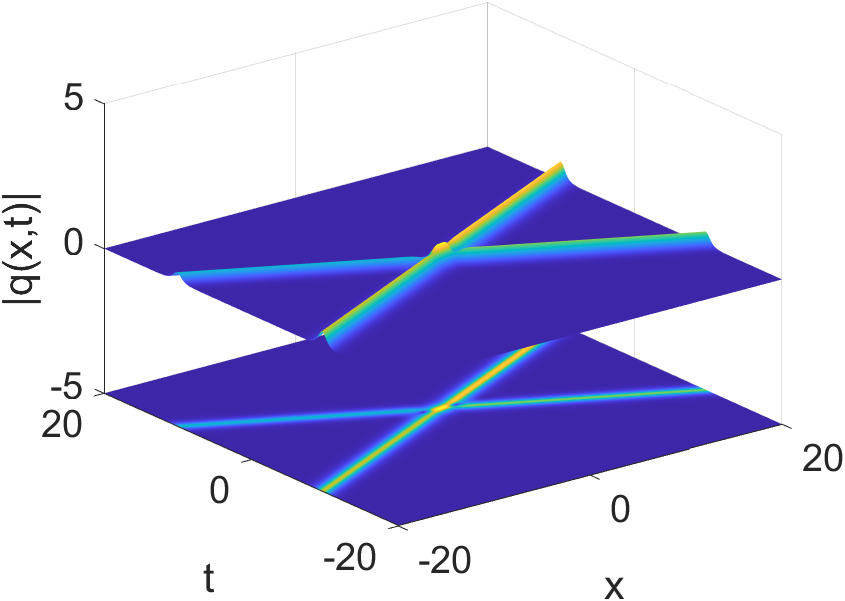}}}
~~{\rotatebox{0}{\includegraphics[width=5.4cm,height=4.2cm,angle=0]{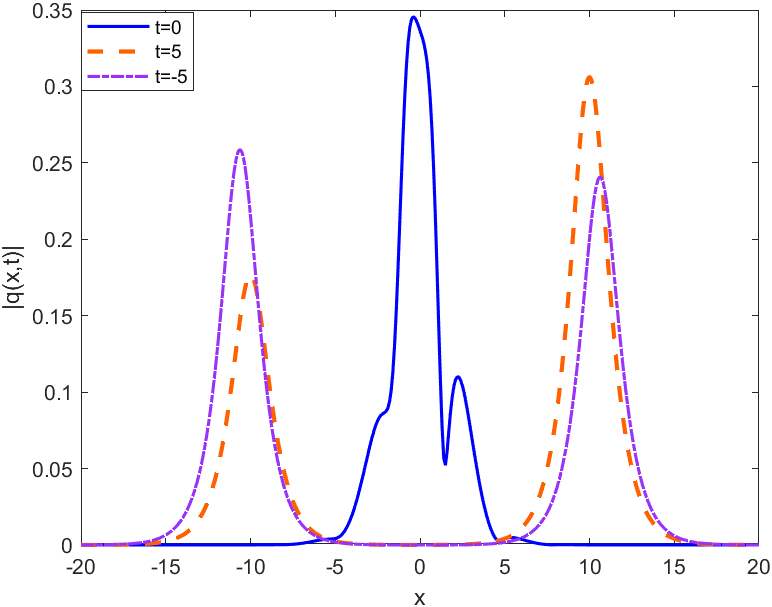}}}
\\
$~~~~~~~~~~~~~~~~~~~~~~~~~~~\textbf{(a)}
~~~~~~~~~~~~~~~~~~~~~~~~~~~~~~~~~~~~~~~~~~~~~~~\textbf{(b)}
$
\\

{\small \textbf{Figure 4.} The 1-soliton solution $q(x,t)$ in \eqref{af-53} with $k_{1}=0.5+0.5\textrm{i}$:
(a) 3D and projection profiles;
(b) wave propagation along the $x$ axis at $t=0$ and $t=\pm5$.}
\vspace{0.2cm}

{\rotatebox{0}{\includegraphics[width=6.2cm,height=4.2cm,angle=0]{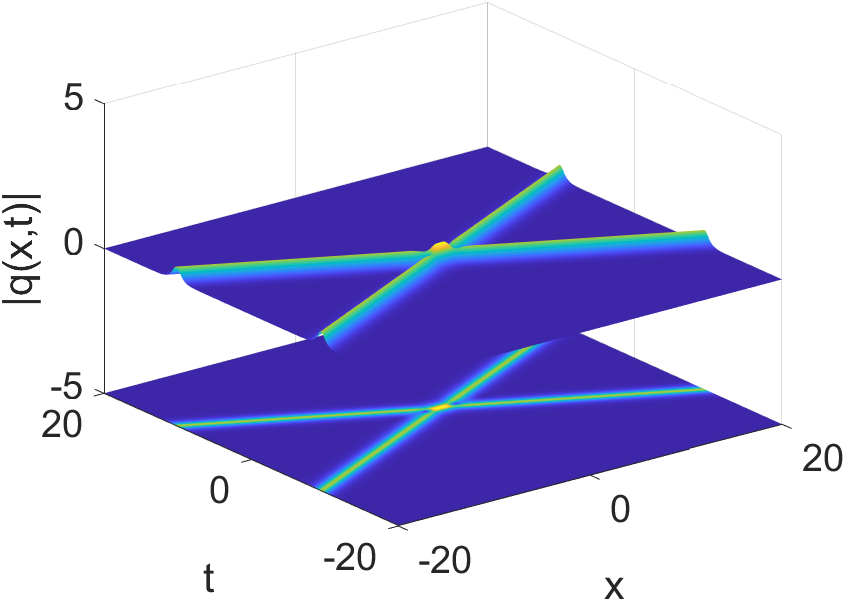}}}
~~{\rotatebox{0}{\includegraphics[width=5.4cm,height=4.2cm,angle=0]{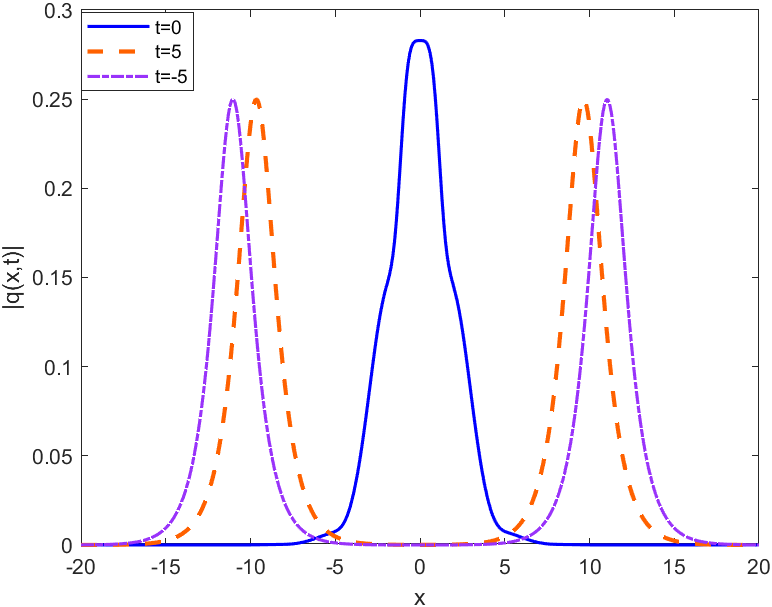}}}
\\
$~~~~~~~~~~~~~~~~~~~~~~~~~~~\textbf{(a)}
~~~~~~~~~~~~~~~~~~~~~~~~~~~~~~~~~~~~~~~~~~~~~~~\textbf{(b)}
$
\\

{\small \textbf{Figure 5.} The 1-soliton solution $q(x,t)$ in \eqref{af-53} with $a_{1}=c_{1}=1$, $k_{1}=0.5+0.5\textrm{i}$:
(a) 3D and projection profiles;
(b) wave propagation along the $x$ axis at $t=0$ and $t=\pm5$.}

\subsection{$2$-soliton dynamics}

In the subsequent, we shall explore the dynamic behaviors of the multi-soliton solution. We choose $N=2$. Because the expansion of the $2$-soliton solution deriving  from Eq. \eqref{af-46} is too tedious, we omit its explicit expression here.

By selecting the appropriate parameter values, the dynamic figures of $2$-soliton solution are exhibited. Moreover, we give the corresponding discussions below.
\vspace{0.2cm}

{\rotatebox{0}{\includegraphics[width=6.2cm,height=4.2cm,angle=0]{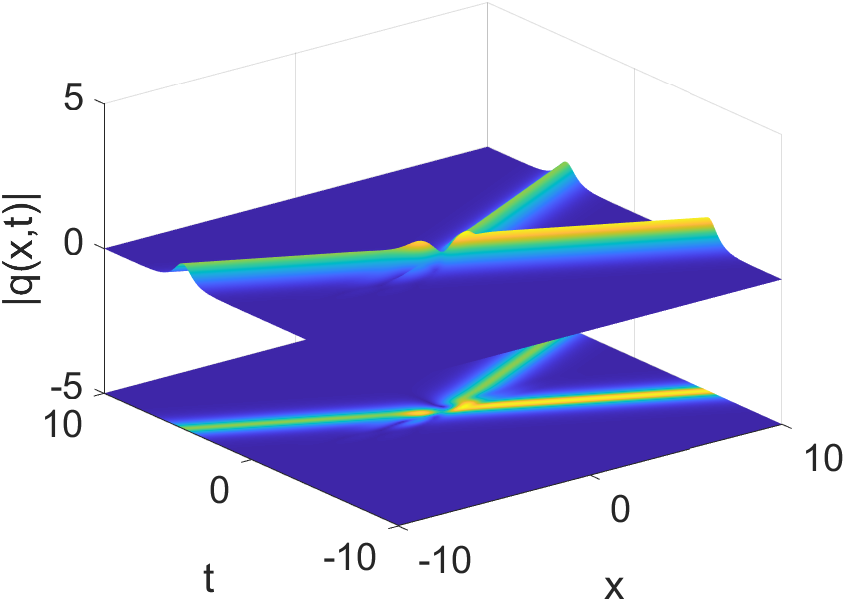}}}
~~{\rotatebox{0}{\includegraphics[width=5.4cm,height=4.2cm,angle=0]{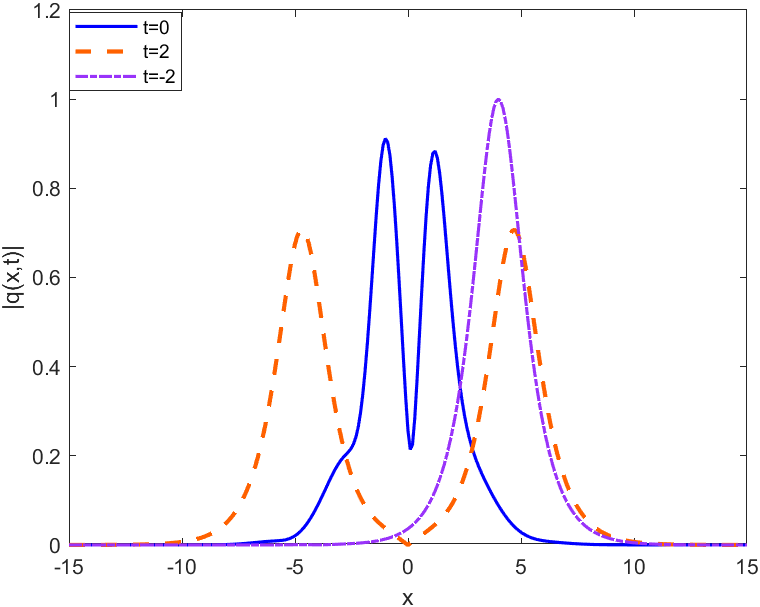}}}
\\
$~~~~~~~~~~~~~~~~~~~~~~~~~~~\textbf{(a)}
~~~~~~~~~~~~~~~~~~~~~~~~~~~~~~~~~~~~~~~~~~~~~~~\textbf{(b)}
$
\\

{\small \textbf{Figure 6.} The interaction of a $2$-soliton in \eqref{af-53} with $k_{1}=0.5+0.5\textrm{i}$ and $k_{2}=-0.5-0.5\textrm{i}$:
(a) 3D and projection profiles;
(b) wave propagation along the $x$ axis at $t=0$ and $t=\pm2$.}
\vspace{0.2cm}

{\rotatebox{0}{\includegraphics[width=6.2cm,height=4.2cm,angle=0]{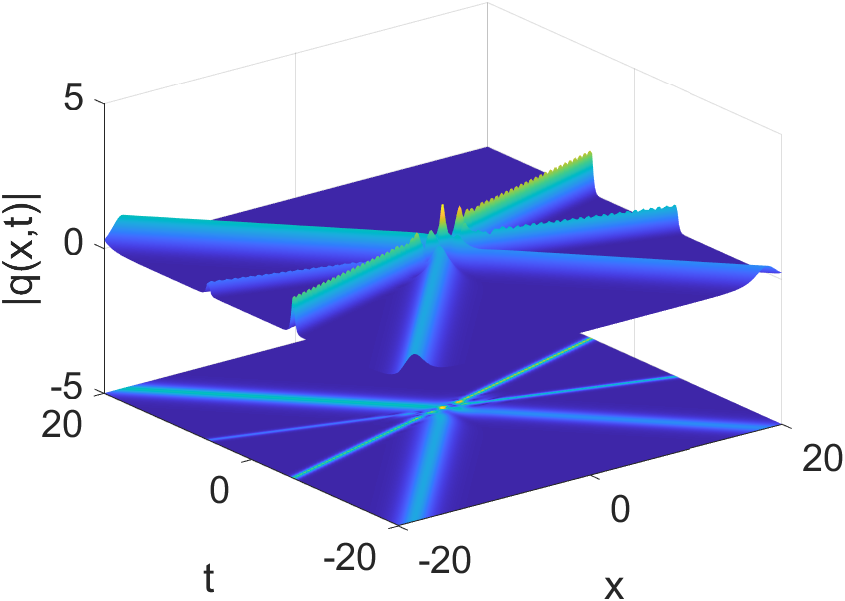}}}
~~{\rotatebox{0}{\includegraphics[width=5.4cm,height=4.2cm,angle=0]{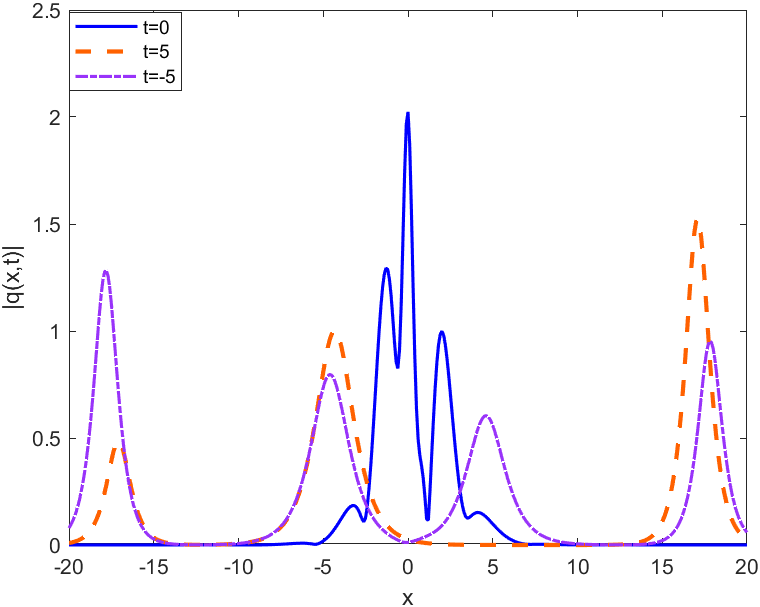}}}
\\
$~~~~~~~~~~~~~~~~~~~~~~~~~~~\textbf{(a)}
~~~~~~~~~~~~~~~~~~~~~~~~~~~~~~~~~~~~~~~~~~~~~~~\textbf{(b)}
$
\\

{\small \textbf{Figure 7.} The interaction between $2$-soliton and $2$-breather in \eqref{af-53} with
$k_{1}=0.4+0.5\textrm{i}$ and $k_{2}=0.7+0.8\textrm{i}$:
(a) 3D and projection profiles;
(b) wave propagation along the $x$ axis at $t=0$ and $t=\pm5$.}
\vspace{0.2cm}

{\rotatebox{0}{\includegraphics[width=6.2cm,height=4.2cm,angle=0]{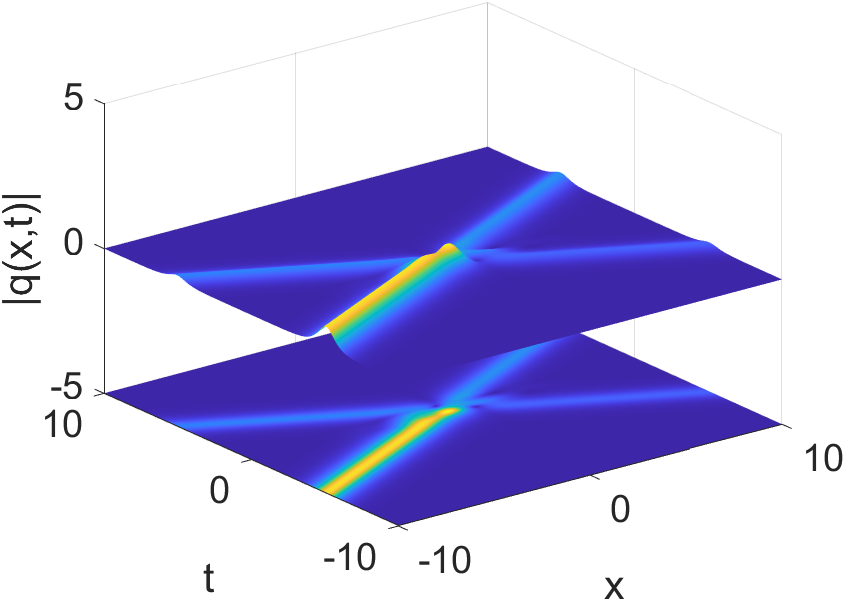}}}
~~{\rotatebox{0}{\includegraphics[width=5.4cm,height=4.2cm,angle=0]{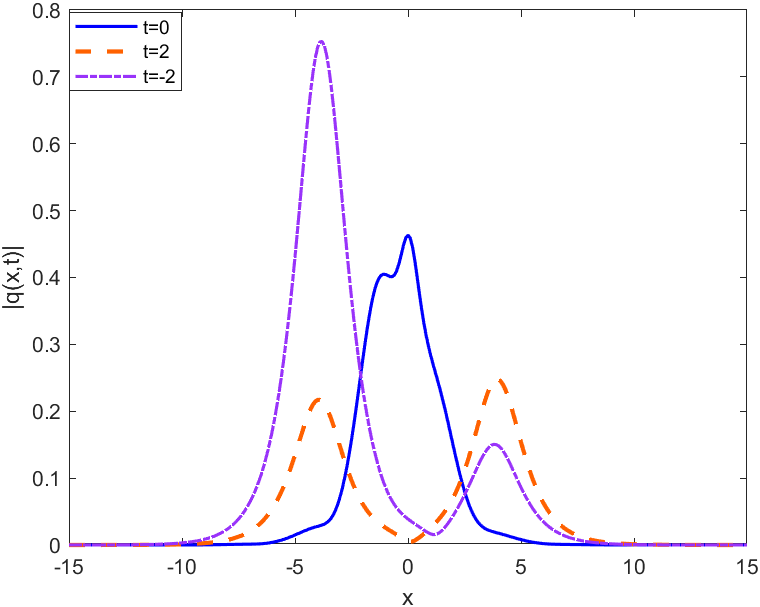}}}
\\
$~~~~~~~~~~~~~~~~~~~~~~~~~~~\textbf{(a)}
~~~~~~~~~~~~~~~~~~~~~~~~~~~~~~~~~~~~~~~~~~~~~~~\textbf{(b)}
$
\\

{\small \textbf{Figure 8.} The interaction of a $2$-soliton in \eqref{af-53} with $a_{1}=b_{1}=c_{1}=d_{1}=1$, $a_{2}=1$, $b_{2}=0$, $c_{2}=2$, $d_{2}=0$, $k_{1}=0.5+0.5\textrm{i}$ and $k_{2}=-0.5-0.5\textrm{i}$:
(a) 3D and projection profiles;
(b) wave propagation along the $x$ axis at $t=0$ and $t=\pm2$.}
\vspace{0.2cm}

{\rotatebox{0}{\includegraphics[width=6.2cm,height=4.2cm,angle=0]{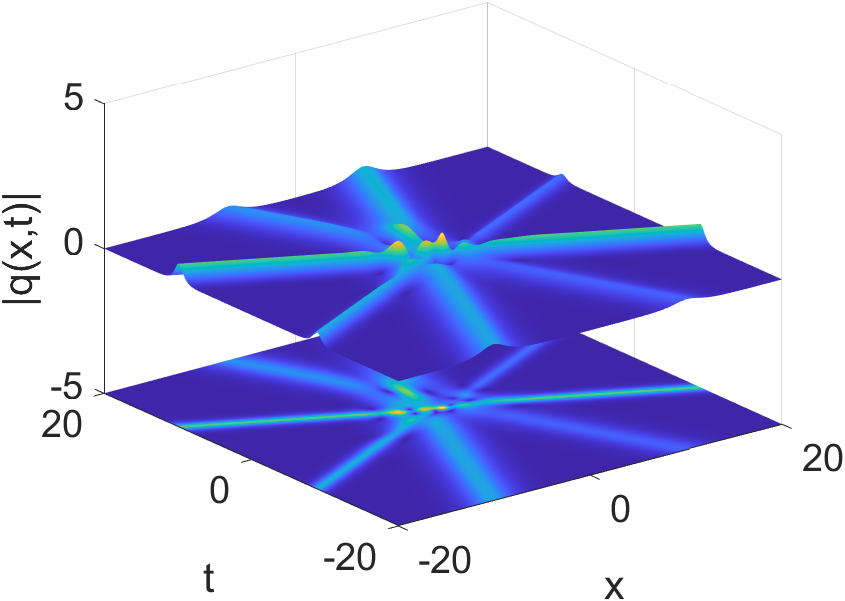}}}
~~{\rotatebox{0}{\includegraphics[width=5.4cm,height=4.2cm,angle=0]{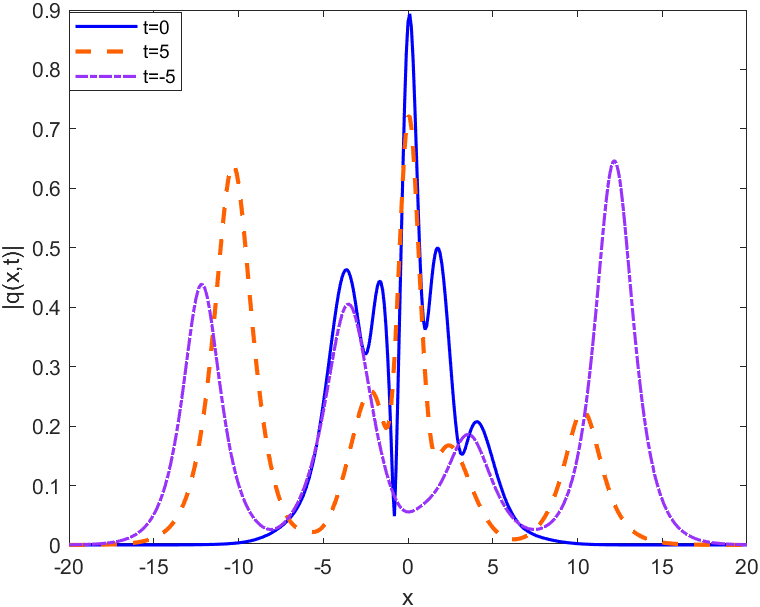}}}
\\
$~~~~~~~~~~~~~~~~~~~~~~~~~~~\textbf{(a)}
~~~~~~~~~~~~~~~~~~~~~~~~~~~~~~~~~~~~~~~~~~~~~~~\textbf{(b)}
$
\\

{\small \textbf{Figure 9.} The interaction between two $2$-soliton in \eqref{af-53} with $a_{1}=b_{1}^{\ast}=1$, $c_{1}=d_{1}^{\ast}=1+\textrm{i}$, $a_{2}=b_{2}^{\ast}=\textrm{i}$, $c_{2}=d_{2}^{\ast}=0.5\textrm{i}$,
$k_{1}=0.3+0.4\textrm{i}$ and $k_{2}=0.5+0.5\textrm{i}$:
(a) 3D and projection profiles;
(b) wave propagation along the $x$ axis at $t=0$ and $t=\pm5$.}
\vspace{0.2cm}

{\rotatebox{0}{\includegraphics[width=6.2cm,height=4.2cm,angle=0]{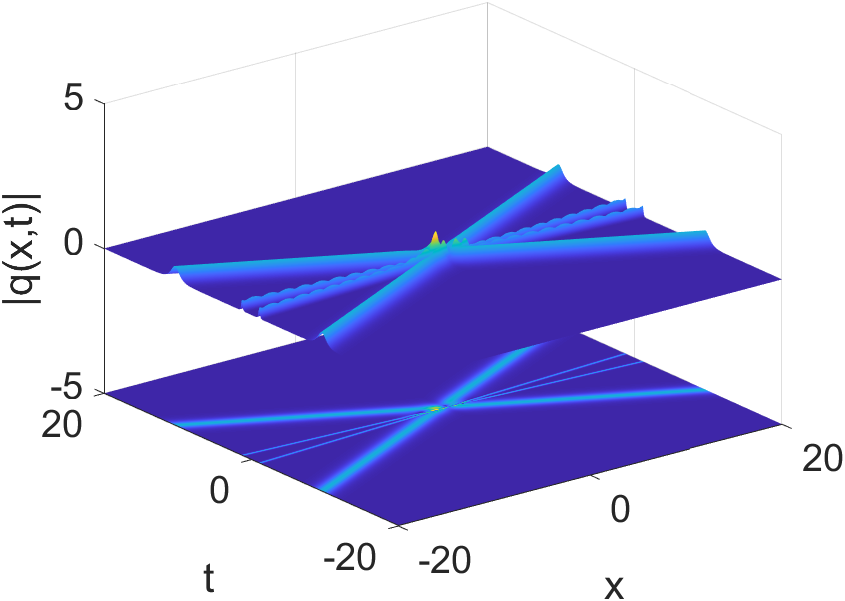}}}
~~{\rotatebox{0}{\includegraphics[width=5.4cm,height=4.2cm,angle=0]{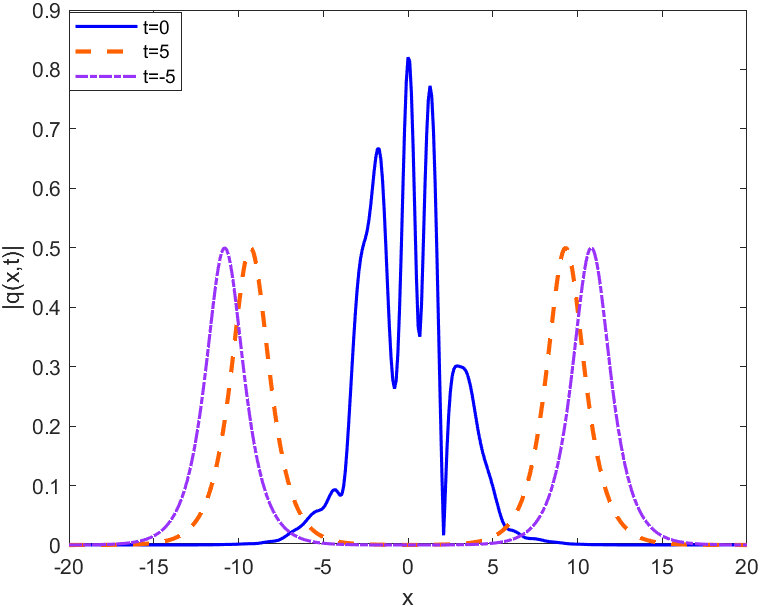}}}
\\
$~~~~~~~~~~~~~~~~~~~~~~~~~~~\textbf{(a)}
~~~~~~~~~~~~~~~~~~~~~~~~~~~~~~~~~~~~~~~~~~~~~~~\textbf{(b)}
$
\\

{\small \textbf{Figure 10.} The interaction between $2$-soliton and $2$-breather in \eqref{af-53} with $a_{1}=b_{1}^{\ast}=\sqrt{2}$, $c_{1}=d_{1}^{\ast}=\sqrt{2}\textrm{i}$, $a_{2}=b_{2}^{\ast}=\textrm{i}$, $c_{2}=d_{2}^{\ast}=1$,
$k_{1}=1.2+0.4\textrm{i}$ and $k_{2}=0.5+0.5\textrm{i}$:
(a) 3D and projection profiles;
(b) wave propagation along the $x$ axis at $t=0$ and $t=\pm5$.}
\vspace{0.2cm}

When $a_{1}=1$, $b_{1}=0$, $c_{1}=0$, $d_{1}=0$, $a_{2}=1$, $b_{2}=0$, $c_{2}=1$ and $d_{2}=0$, we draw Figs. 6 and 7. It follows from Fig. 6 that we see a $2$-soliton interaction, where a soliton disappears after the collision. Fig. 7 displays a $2$-breather and a $2$-soliton interaction, where a soliton disappears after the collision. Compared Figs. 6 with 8, we learn that the value of parameter $c_{1}$ determines whether the soliton disappears after the collision. In other words, the amplitude of a soliton will directly reduce to zero after the collision if $c_{1}=0$ or $c_{2}=0$. Fig. 9 depicts two $2$-soliton interaction. Fig. 10 shows the interaction between a $2$-soliton and a $2$-breather. From the pattern of the wave propagation along the $x$ axis at $t=\pm5$ in Fig. 10(b), we easily observe that when $|a_{1}|=|c_{1}|$ and $|a_{2}|=|c_{2}|$, they keep travelling in their original directions and amplitudes after the collision. This is clearly different from what are exhibited in Figs. 6(b)-9(b). Additionally, with all other conditions constant, let $k_{1}=-k_{2}$, we can get a simple interaction between $N$-soliton.

\section{Inverse scattering transform with high-order poles}

Recall the process of the case of simple pole, if $\det\mathcal{P}_{1}$ has a number of $2N$ high-order zeros $\{k_{i}\}_{1}^{2N}$ in $D_{+}$, according to the symmetry reductions Eqs. \eqref{af-24} and \eqref{af-27}, then $k_{N+i}=-k_{i}$, $i=1,2,\ldots,N$, and there are also $2N$ high-order zeros $\widehat{k}_{j}=k_{j}^{\ast}$, $j=1,2,\ldots,2N$ in $D_{-}$. The geometric multiplication of these high-order zeros equals to 1, and the corresponding order of a zero $k_{i}$ is $R[i]$ with $R=\left(n_{1},n_{2},\ldots,n_{2N}\right)$. It is noted that $n_{N+i}=n_{i}$, $i=1,2,\ldots,N$, so the algebraic multiplication reads $2N_{0}$, in which $N_{0}=\sum_{i=1}^{N}n_{i}$.

Based on the Proposition 2.3 and Proposition 2.5, we have
\begin{align*}
s_{55}(k)&=\prod_{i=1}^{2N}(k-k_{i})^{n_{i}}s_{55}^{[0]}(k),\notag\\
r_{55}(k)&=\prod_{i=1}^{2N}(k-\widehat{k}_{i})^{n_{i}}r_{55}^{[0]}(k),
\end{align*}
where the initial scattering data $s_{55}^{[0]}(k)$ and $r_{55}^{[0]}(k)$ both are not zeros.

In Refs. \cite{Bian-2015,Bo-2019,Bo-2018,Min-2023}, authors dealt with high-order zeros by means of a simple limiting process. Therefore, in order to obtain the general high-order soliton formula for the ngSS equation \eqref{ngss}, it is natural to add the perturbation parameters $\varepsilon$ and $\hat{\varepsilon}$ to the discrete spectrum $\{k_{l},\widehat{k}_{l}\}$, $l=1,2,\ldots,2N$, respectively.

To be concrete, we have a new discrete spectrum $\{k_{l}+\varepsilon,\widehat{k}_{l}+\widehat{\varepsilon}\}$, $l=1,2,\ldots,2N$. The corresponding perturbation eigenvectors \eqref{af-32} become
\begin{align}\label{bf-2}
&U_{l}(\varepsilon)= \left\{\begin{array}{cc}
                 \textrm{e}^{\theta_{l}(x,t,\varepsilon)\sigma_{3}}U_{l,0}(\varepsilon), & 1\leq l\leq N, \\
                 \Lambda\textrm{e}^{\theta_{l-N}(-x,t,\varepsilon)\sigma_{3}}U_{l-N,0}(\varepsilon), & N+1\leq l\leq 2N,
               \end{array}\right.\notag\\
&\widehat{U}_{l}(\widehat{\varepsilon})= \left\{\begin{array}{cc}
                 \widehat{U}_{l,0}(\widehat{\varepsilon})\textrm{e}^{\theta_{l}^{\ast}(x,t,\varepsilon)\sigma_{3}}, & 1\leq l\leq N, \\
                 \widehat{U}_{l-N,0}(\widehat{\varepsilon})\textrm{e}^{\theta_{l-N}^{\ast}(-x,t,\varepsilon)\sigma_{3}}\Lambda, & N+1\leq l\leq 2N,
               \end{array}\right.
\end{align}
with
\begin{equation}\label{bf-3}
\theta_{l}(x,t,\varepsilon):=\theta(x,t,k_{l}+\varepsilon)=\theta_{l}(x,t)+\left(\textrm{i}x
+12\textrm{i}k_{l}^{2}t\right)\varepsilon+12\textrm{i}k_{l}t\varepsilon^{2}
+4\textrm{i}t\varepsilon^{3},
\end{equation}
where $\theta_{l}(x,t):=\theta(x,t,k_{l})$ is given by \eqref{af-20}.

We introduce the initial vectors
\begin{equation*}
U_{l,0}(\varepsilon)=\left(A_{l},B_{l},C_{l},D_{l},1\right)^{\mathrm{T}},~~
\widehat{U}_{l,0}(\widehat{\varepsilon})=\left(\widehat{A}_{l},\widehat{B}_{l},
\widehat{C}_{l},\widehat{D}_{l},1\right),~~1\leq l\leq N,
\end{equation*}
and their perturbation expressions are
\begin{align}\label{bf-4}
&A_{l}=\textrm{e}^{\sum_{i=0}^{+\infty}a_{l}^{[i]}\varepsilon^{i}},
~~B_{l}=\textrm{e}^{\sum_{i=0}^{+\infty}b_{l}^{[i]}\varepsilon^{i}},
~~C_{l}=\textrm{e}^{\sum_{i=0}^{+\infty}c_{l}^{[i]}\varepsilon^{i}},
~~D_{l}=\textrm{e}^{\sum_{i=0}^{+\infty}d_{l}^{[i]}\varepsilon^{i}},\notag\\
&\widehat{A}_{l}=\textrm{e}^{\sum_{i=0}^{+\infty}\widehat{a}_{l}^{[i]}\widehat{\varepsilon}^{i}},
~~\widehat{B}_{l}=\textrm{e}^{\sum_{i=0}^{+\infty}\widehat{b}_{l}^{[i]}\widehat{\varepsilon}^{i}},
~~\widehat{C}_{l}=\textrm{e}^{\sum_{i=0}^{+\infty}\widehat{c}_{l}^{[i]}\widehat{\varepsilon}^{i}},
~~\widehat{D}_{l}=\textrm{e}^{\sum_{i=0}^{+\infty}\widehat{d}_{l}^{[i]}\widehat{\varepsilon}^{i}},
\end{align}
where parameters $a_{l}^{[i]}$, $b_{l}^{[i]}$, $c_{l}^{[i]}$, $d_{l}^{[i]}$, $\widehat{a}_{l}^{[i]}$, $\widehat{b}_{l}^{[i]}$, $\widehat{c}_{l}^{[i]}$ and $\widehat{d}_{l}^{[i]}$ are free complex constants.

Expanding the perturbation eigenvectors defined by \eqref{bf-2} at $(\varepsilon,\widehat{\varepsilon})=(0,0)$ yields that
\begin{equation}\label{bf-5}
U_{l}(\varepsilon)=\sum_{i=0}^{+\infty}U_{l}^{[i]}\varepsilon^{i},~~
\widehat{U}_{l}(\widehat{\varepsilon})=\sum_{i=0}^{+\infty}\widehat{U}_{l}^{[i]}
\widehat{\varepsilon}^{i}.
\end{equation}
Furthermore,
\begin{equation}\label{bf-6}
\frac{\widehat{U}_{i}(\widehat{\varepsilon})U_{j}(\varepsilon)}
{k_{j}+\varepsilon-(\widehat{k}_{i}+\widehat{\varepsilon})}=
\sum_{l_{1}=0}^{+\infty}\sum_{l_{2}=0}^{+\infty}M_{j,i}^{[l_{2},l_{1}]}\varepsilon^{l_{2}}
\widehat{\varepsilon}^{l_{1}},~~1\leq i,j\leq 2N.
\end{equation}
Inserting expressions \eqref{bf-3}, \eqref{bf-4}, \eqref{bf-5} and \eqref{bf-6} into the potential \eqref{af-40}, and taking limit of the perturbation parameters $(\varepsilon,\widehat{\varepsilon})\rightarrow(0,0)$, we conclude the following theorem.
\vspace{0.2cm}

\noindent
\textbf{Theorem 4.1} Let $n_{i}$ denote the corresponding order of $2N$ high-order zeros, in which $n_{N+i}=n_{i}$, $i=1,2,\ldots,N$, and $N_{0}=\sum_{i=1}^{N}n_{i}$. The $N_{0}$-th high-order $N$-soliton solution of the ngSS equation \eqref{ngss} can be given by
\begin{align}\label{bf-7}
q(x,t)=2\textrm{i}\frac{\det \mathcal{H}}{\det \mathcal{M}},
\end{align}
with
\begin{equation}\nonumber
\mathcal{H}=\left(
    \begin{array}{cc}
      \mathcal{M} & \widehat{\chi}_{5} \\
      \chi_{1} & 0 \\
    \end{array}
  \right),~~\mathcal{M}=\left(M_{j,i}^{[l_{2},l_{1}]}\right)_{2N_{0}\times 2N_{0}},
\end{equation}
where $M_{j,i}^{[l_{2},l_{1}]}$ is defined by \eqref{bf-6}, $1\leq i,j\leq 2N$, $0\leq l_{1}\leq R[i]$, $0\leq l_{2}\leq R[j]$, moreover, $\widehat{\chi}_{5}$ and $\chi_{1}$ severally denote the fifth column and the first row of the vectors $\widehat{\chi}$ and $\chi$,
\begin{equation}\nonumber
\begin{aligned}
\widehat{\chi}=&\left(
                    \widehat{U}_{1}^{[0]}, \ldots, \widehat{U}_{1}^{[n_{1}-1]}, \ldots, \widehat{U}_{N}^{[0]}, \ldots, \widehat{U}_{N}^{[n_{N}-1]},\right.\notag\\
                    &\left.
                    \widehat{U}_{N+1}^{[0]}, \ldots, \widehat{U}_{N+1}^{[n_{1}-1]}, \ldots, \widehat{U}_{2N}^{[0]}, \ldots, \widehat{U}_{2N}^{[n_{N}-1]}
                \right)^{\mathrm{T}}_{2N_{0}\times5},\notag\\
\chi=&\left(
                    U_{1}^{[0]}, \ldots, U_{1}^{[n_{1}-1]}, \ldots, U_{N}^{[0]}, \ldots, U_{N}^{[n_{N}-1]}, \right.\notag\\
                    &\left.
                    U_{N+1}^{[0]}, \ldots, U_{N+1}^{[n_{1}-1]}, \ldots, U_{2N}^{[0]}, \ldots, U_{2N}^{[n_{N}-1]}
                \right)_{5\times 2N_{0}}.
\end{aligned}
\end{equation}

\noindent
\textbf{Remark 4.2} When $n_{l}=1$, $l=1,2,\ldots,2N$, the $N_{0}$-th high-order $N$-soliton solution \eqref{bf-7} reduces to the case of simple $N$-soliton solution.
\vspace{0.2cm}

For the purpose of exhibiting the dynamic behavior of the $N_{0}$-th high-order $N$-soliton solution \eqref{bf-7} more intuitively, without loss of generality, we discuss the second-order $1$-soliton and the third-order $2$-soliton solutions in detail in the following sections.

\subsection{Second-order $1$-soliton solution}
We set $N=1$ and $N_{0}=2$, that is $n_{1}=2$ and $n_{l}=0$, $l=2,\ldots,N$, moreover, $a_{1}^{[k]}=b_{1}^{[k]}=c_{1}^{[k]}=d_{1}^{[k]}=\widehat{a}_{1}^{[k]}=\widehat{b}_{1}^{[k]}
=\widehat{c}_{1}^{[k]}=\widehat{d}_{1}^{[k]}=0$, $k\geq1$, then derive the expression of the second-order $1$-soliton solution with the form
\begin{align}\label{bf-8}
q(x,t)=2\textrm{i}\frac{\det \mathcal{H}}{\det \mathcal{M}},
\end{align}
with
\begin{equation}\nonumber
\mathcal{H}=\left(
    \begin{array}{cc}
      \mathcal{M} & \widehat{\chi}_{5} \\
      \chi_{1} & 0 \\
    \end{array}
  \right),~~\mathcal{M}=\left(M_{j,i}^{[l_{2},l_{1}]}\right)_{4\times 4},~~i,j=1,2.
\end{equation}
It follows from Eq. \eqref{bf-6} that we have
\begin{equation}\nonumber
M_{j,i}^{[l_{2},l_{1}]}=\mathop{\lim}\limits_{\varepsilon,\widehat{\varepsilon}\rightarrow0}\frac{1}{
(l_{2}-1)!(l_{1}-1)}
\frac{\partial^{l_{2}-1}\cdot\partial^{l_{1}-1}}{\partial\widehat{\varepsilon}^{l_{2}-1}
\cdot\partial\varepsilon^{l_{1}-1}}\left[\frac{\widehat{U}_{i}(\widehat{\varepsilon})U_{j}(\varepsilon)}
{k_{j}+\varepsilon-(\widehat{k}_{i}+\widehat{\varepsilon})}\right],~~l_{1}, l_{2}=1,2.
\end{equation}
According to the Taylor expansion formulae, we calculate that
\begin{equation}\nonumber
\begin{aligned}
\widehat{\chi}_{5}=&\left(
                      \begin{array}{cccc}
                        1-\theta_{1}^{\ast}(x,t) & \textrm{i}x+12\textrm{i}(k_{1}^{\ast})^{2}t &
                        -1+\theta_{1}^{\ast}(-x,t) &
                        \textrm{i}x-12\textrm{i}(k_{1}^{\ast})^{2}t \\
                      \end{array}
                    \right)^{\mathrm{T}},\notag\\
\chi_{1}=&\left(
    \begin{array}{cccc}
    1+a_{1}^{[0]}+\theta_{1}(x,t) &
    \textrm{i}x+12\textrm{i}k_{1}^{2}t &
    1+c_{1}^{[0]}+\theta_{1}(-x,t) &
    -\textrm{i}x+12\textrm{i}k_{1}^{2}t\\
    \end{array}
    \right).
\end{aligned}
\end{equation}
As a result, we display dynamic behaviors of the second-order $1$-soliton solution \eqref{bf-8} by selecting appropriate parameters values.
\vspace{0.2cm}

{\rotatebox{0}{\includegraphics[width=6.2cm,height=4.2cm,angle=0]{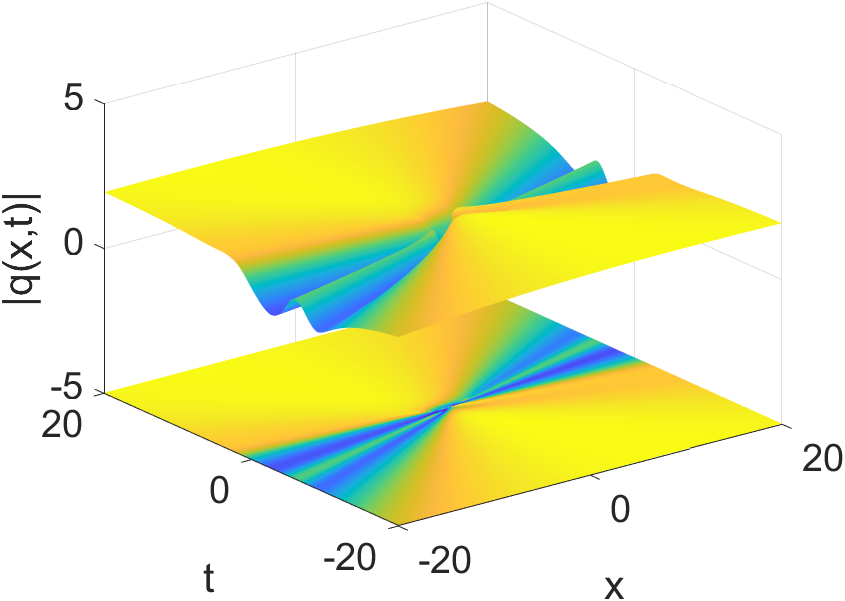}}}
~~{\rotatebox{0}{\includegraphics[width=5.4cm,height=4.2cm,angle=0]{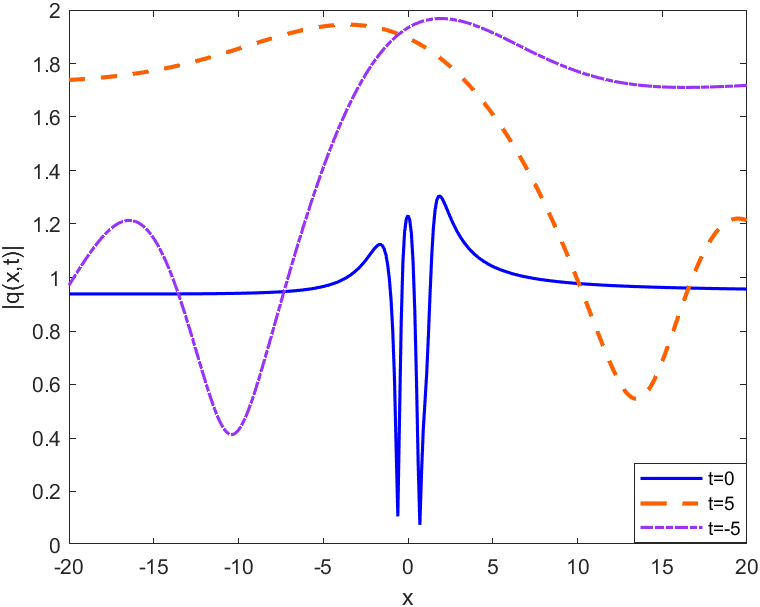}}}
\\
$~~~~~~~~~~~~~~~~~~~~~~~~~~~\textbf{(a)}
~~~~~~~~~~~~~~~~~~~~~~~~~~~~~~~~~~~~~~~~~~~~~~~\textbf{(b)}
$
\\

{\small \textbf{Figure 11.} The second-order 1-soliton solution $q(x,t)$ in \eqref{bf-8} with $a_{1}^{[0]}=b_{1}^{[0]}=c_{1}^{[0]}=d_{1}^{[0]}=
\widehat{a}_{1}^{[0]}=\widehat{b}_{1}^{[0]}=\widehat{c}_{1}^{[0]}=\widehat{d}_{1}^{[0]}=0$ and $k_{1}=0.3+\textrm{i}$:
(a) 3D and projection profiles;
(b) wave propagation along the $x$ axis at $t=0$ and $t=\pm5$.}
\vspace{0.2cm}

{\rotatebox{0}{\includegraphics[width=6.2cm,height=4.2cm,angle=0]{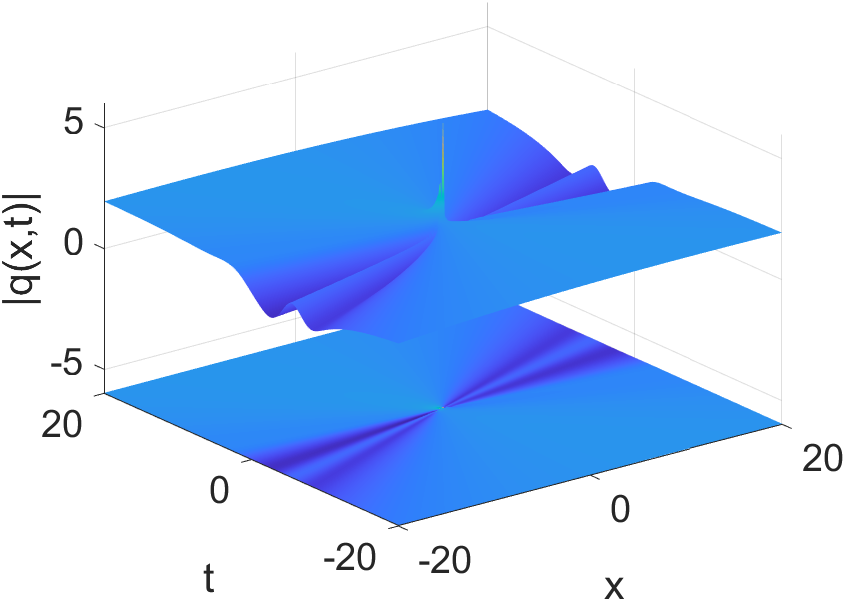}}}
~~{\rotatebox{0}{\includegraphics[width=5.4cm,height=4.2cm,angle=0]{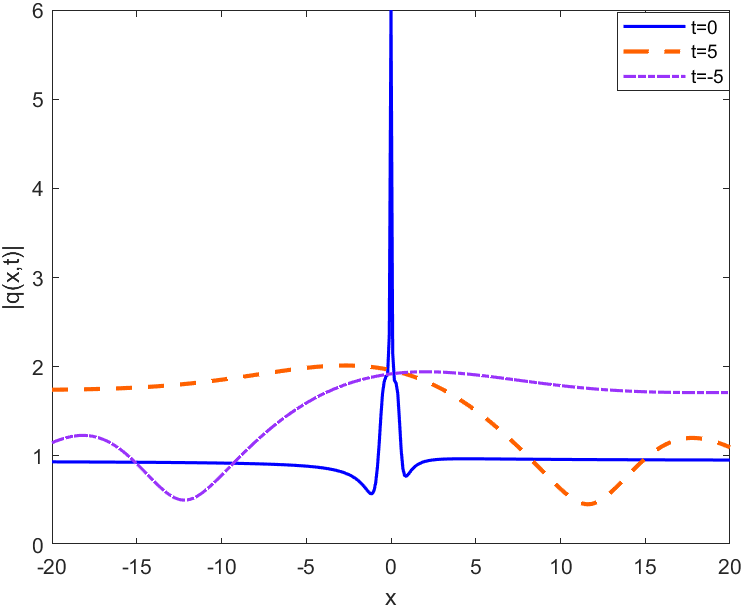}}}
\\
$~~~~~~~~~~~~~~~~~~~~~~~~~~~\textbf{(a)}
~~~~~~~~~~~~~~~~~~~~~~~~~~~~~~~~~~~~~~~~~~~~~~~\textbf{(b)}
$
\\

{\small \textbf{Figure 12.} The second-order 1-soliton solution $q(x,t)$ in \eqref{bf-8} with $a_{1}^{[0]}=b_{1}^{[0]}=c_{1}^{[0]}=d_{1}^{[0]}=
\widehat{a}_{1}^{[0]}=\widehat{b}_{1}^{[0]}=\widehat{c}_{1}^{[0]}=\widehat{d}_{1}^{[0]}=-1.5$ and $k_{1}=0.3+\textrm{i}$:
(a) 3D and projection profiles;
(b) wave propagation along the $x$ axis at $t=0$ and $t=\pm5$.}
\vspace{0.2cm}

{\rotatebox{0}{\includegraphics[width=6.2cm,height=4.2cm,angle=0]{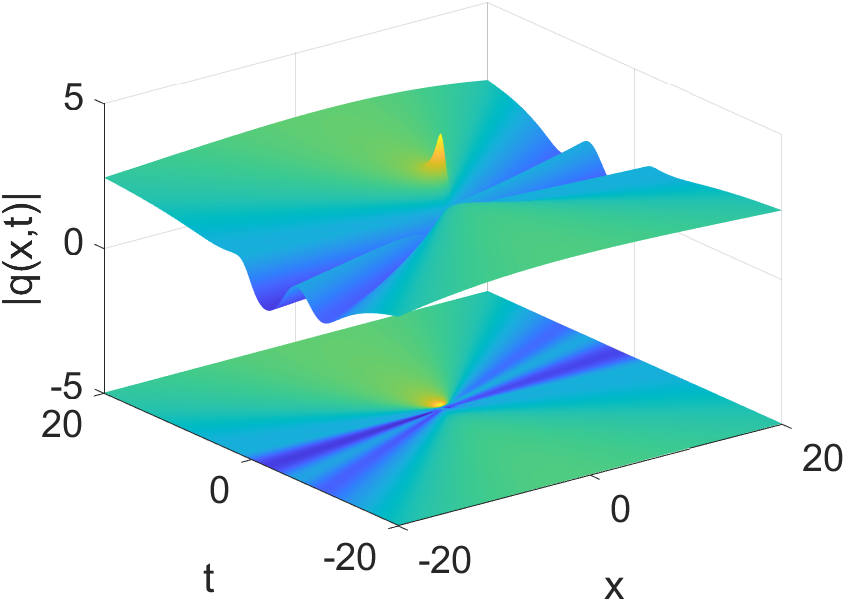}}}
~~{\rotatebox{0}{\includegraphics[width=5.4cm,height=4.2cm,angle=0]{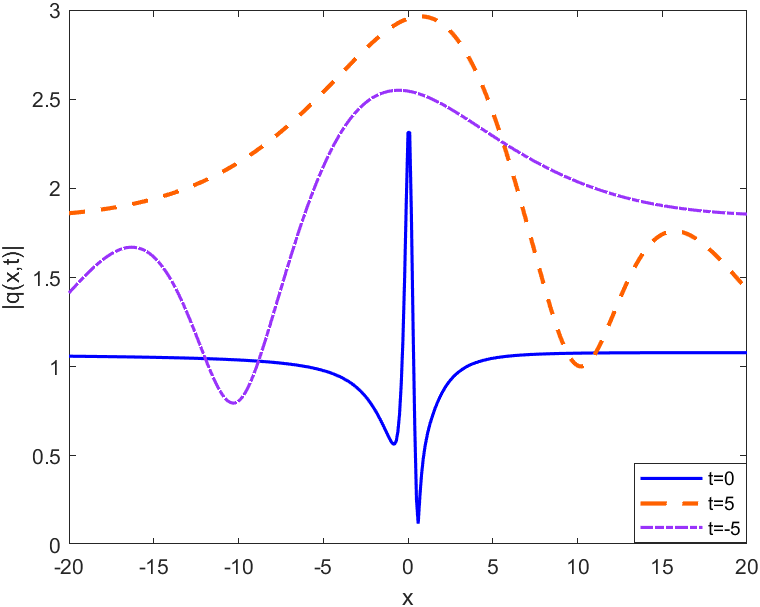}}}
\\
$~~~~~~~~~~~~~~~~~~~~~~~~~~~\textbf{(a)}
~~~~~~~~~~~~~~~~~~~~~~~~~~~~~~~~~~~~~~~~~~~~~~~\textbf{(b)}
$
\\

{\small \textbf{Figure 13.} The second-order 1-soliton solution $q(x,t)$ in \eqref{bf-8}:
(a) 3D and projection profiles;
(b) wave propagation along the $x$ axis at $t=0$ and $t=\pm5$.}
\vspace{0.2cm}

By keeping the initial value $k_{1}$ unchange, the number of peaks of the second-order $1$-soliton solution depends on the values of parameters $a_{1}^{[0]}$, $b_{1}^{[0]}$, $c_{1}^{[0]}$, $d_{1}^{[0]}$, $\widehat{a}_{1}^{[0]}$, $\widehat{b}_{1}^{[0]}$, $\widehat{c}_{1}^{[0]}$ and $\widehat{d}_{1}^{[0]}$, as shown in Figs. 11 and 12. Taking $a_{1}^{[0]}=c_{1}^{[0]}=-1+\textrm{i}$, $\widehat{a}_{1}^{[0]}=\widehat{c}_{1}^{[0]}=-1-\textrm{i}$, $b_{1}^{[0]}=d_{1}^{[0]}= \widehat{b}_{1}^{[0]}=\widehat{d}_{1}^{[0]}=-1$ and $k_{1}=0.4+\textrm{i}$, the 3D, projection profiles and wave propagation along the $x$
axis at $t = 0$ and $t =\pm5$ is drawn in Fig. 13. The second-order $1$-soliton solution in above figures show more complicated wave and trajectory structures compared with the simple case.

\subsection{Third-order $2$-soliton solution}

In this section, we set $N=2$, $n_{1}=2$, $n_{2}=1$ and $n_{l}=0$, $l=3,\ldots,N$, then calculate $N_{0}=3$. In addition, taking $a_{1}^{[k]}=b_{1}^{[k]}=c_{1}^{[k]}=d_{1}^{[k]}=\widehat{a}_{1}^{[k]}=\widehat{b}_{1}^{[k]}
=\widehat{c}_{1}^{[k]}=\widehat{d}_{1}^{[k]}=0$, $k\geq2$, the expression of the third-order $2$-soliton solution is derived, whose form is
\begin{align}\label{bf-9}
q(x,t)=2\textrm{i}\frac{\det \mathcal{H}}{\det \mathcal{M}},
\end{align}
with
\begin{equation}\nonumber
\mathcal{H}=\left(
    \begin{array}{cc}
      \mathcal{M} & \widehat{\chi}_{5} \\
      \chi_{1} & 0 \\
    \end{array}
  \right),~~\mathcal{M}=\left(M_{j,i}^{[l_{2},l_{1}]}\right)_{6\times 6},~~i,j=1,2,3,4.
\end{equation}
It follows from Eq. \eqref{bf-6} that we have
\begin{equation}\nonumber
\begin{aligned}
M_{j,i}^{[l_{2},l_{1}]}=&\mathop{\lim}\limits_{\varepsilon,\widehat{\varepsilon}\rightarrow0}\frac{1}{
(l_{2}-1)!(l_{1}-1)}
\frac{\partial^{l_{2}-1}\cdot\partial^{l_{1}-1}}{\partial\widehat{\varepsilon}^{l_{2}-1}
\cdot\partial\varepsilon^{l_{1}-1}}\left[\frac{\widehat{U}_{i}(\widehat{\varepsilon})U_{j}(\varepsilon)}
{k_{j}+\varepsilon-(\widehat{k}_{i}+\widehat{\varepsilon})}\right],\notag\\
&1\leq l_{1}\leq R[i],~~1\leq l_{2}\leq R[j].
\end{aligned}
\end{equation}
It follows from the Taylor expansion formulae that one arrives at
\begin{equation}\nonumber
\begin{aligned}
\widehat{\chi}_{5}=&\left(\begin{array}{ccc}
                      1-\theta_{1}^{\ast}(x,t) & \textrm{i}x+12\textrm{i}(k_{1}^{\ast})^{2}t & 1-\theta_{2}^{\ast}(x,t)
                    \end{array}\right.\notag\\
                    &\left.\begin{array}{ccc}
                             -1+\theta_{1}^{\ast}(-x,t) & \textrm{i}x-12\textrm{i}(k_{1}^{\ast})^{2}t & -1+\theta_{2}^{\ast}(-x,t)
                           \end{array}
                    \right)^{\mathrm{T}},\notag\\
\chi_{1}=&\left(\begin{array}{ccc}
                  1+a_{1}^{[0]}+\theta_{1}(x,t) & \textrm{i}x+12\textrm{i}k_{1}^{2}t+a_{1}^{[1]} & 1+a_{2}^{[0]}+\theta_{2}(x,t)
                \end{array}\right.\notag\\
&\left.\begin{array}{ccc}
        1+c_{1}^{[0]}+\theta_{1}(-x,t) & -\textrm{i}x+12\textrm{i}k_{1}^{2}t+c_{1}^{[1]} & 1+c_{2}^{[0]}+\theta_{2}(-x,t)
      \end{array}
\right).
\end{aligned}
\end{equation}

Further on, when $a_{1}^{[0]}=b_{1}^{[0]}=c_{1}^{[0]}=d_{1}^{[0]}=
\widehat{a}_{1}^{[0]}=\widehat{b}_{1}^{[0]}=\widehat{c}_{1}^{[0]}=\widehat{d}_{1}^{[0]}
=-1$ and $a_{1}^{[1]}=b_{1}^{[1]}=c_{1}^{[1]}=d_{1}^{[1]}=\widehat{a}_{1}^{[1]}=\widehat{b}_{1}^{[1]}
=\widehat{c}_{1}^{[1]}=\widehat{d}_{1}^{[1]}=0$, we have the following dynamic figures.
\vspace{0.2cm}

{\rotatebox{0}{\includegraphics[width=6.2cm,height=4.2cm,angle=0]{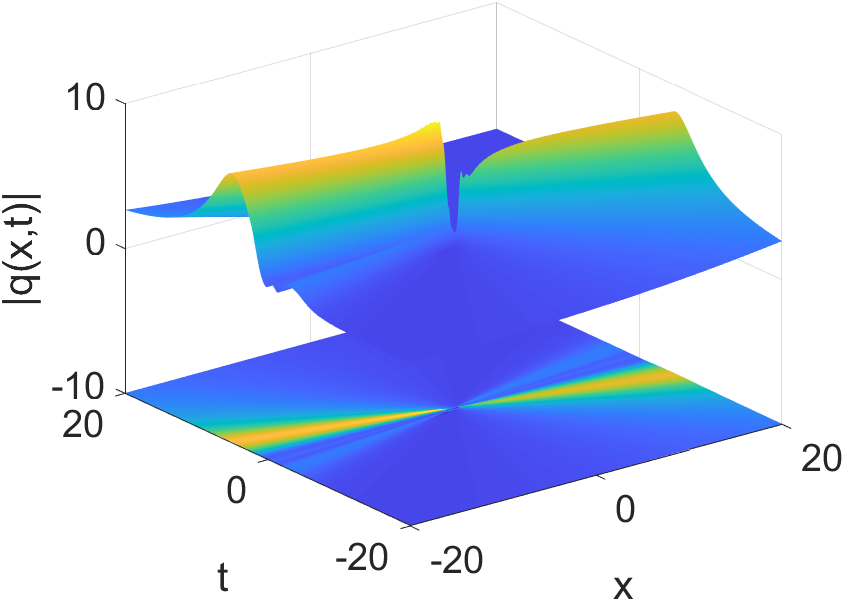}}}
~~{\rotatebox{0}{\includegraphics[width=5.4cm,height=4.2cm,angle=0]{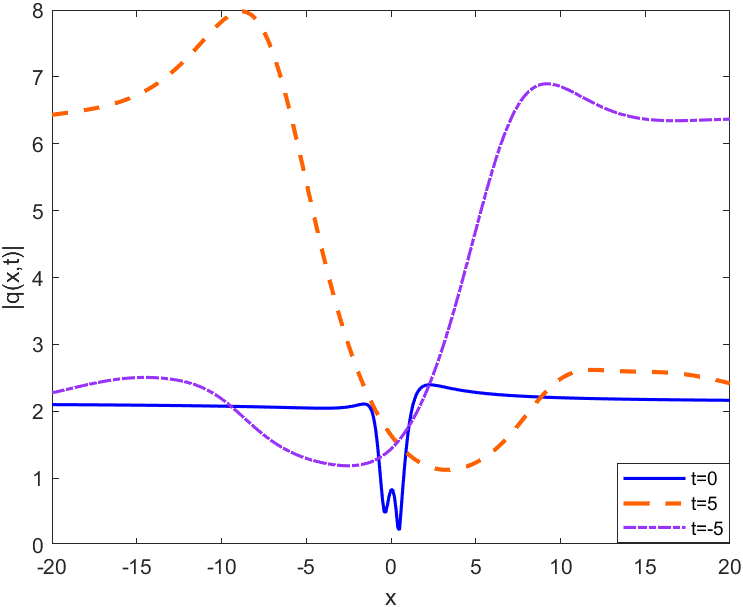}}}
\\
$~~~~~~~~~~~~~~~~~~~~~~~~~~~\textbf{(a)}
~~~~~~~~~~~~~~~~~~~~~~~~~~~~~~~~~~~~~~~~~~~~~~~\textbf{(b)}
$
\\

{\small \textbf{Figure 14.} The third-order 2-soliton solution $q(x,t)$ in \eqref{bf-9} with $a_{2}^{[0]}=b_{2}^{[0]}=c_{2}^{[0]}=d_{2}^{[0]}=0$, $k_{1}=0.3+1\textrm{i}$ and $k_{2}=1+1\textrm{i}$:
(a) 3D and projection profiles;
(b) wave propagation along the $x$ axis at $t=0$ and $t=\pm5$.}
\vspace{0.2cm}

{\rotatebox{0}{\includegraphics[width=6.2cm,height=4.2cm,angle=0]{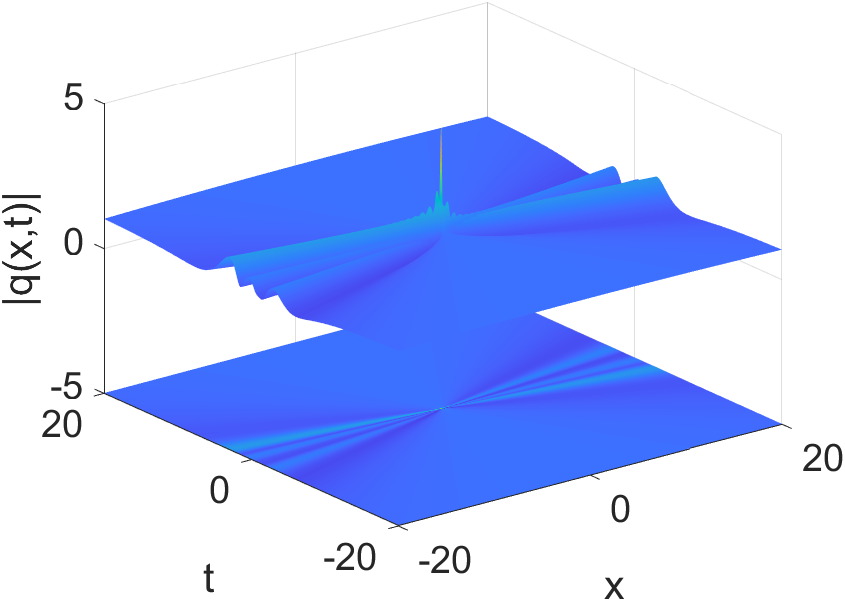}}}
~~{\rotatebox{0}{\includegraphics[width=5.4cm,height=4.2cm,angle=0]{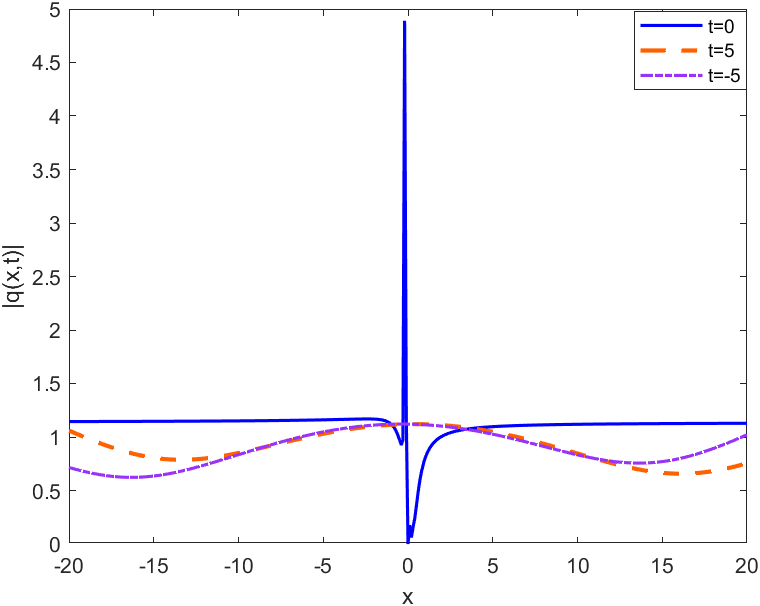}}}
\\
$~~~~~~~~~~~~~~~~~~~~~~~~~~~\textbf{(a)}
~~~~~~~~~~~~~~~~~~~~~~~~~~~~~~~~~~~~~~~~~~~~~~~\textbf{(b)}
$
\\

{\small \textbf{Figure 15.} The third-order 2-soliton solution $q(x,t)$ in \eqref{bf-9} with $a_{2}^{[0]}=b_{2}^{[0]}=c_{2}^{[0]}=d_{2}^{[0]}=-1$, $k_{1}=0.5+1.2\textrm{i}$ and $k_{2}=0.6+1.2\textrm{i}$:
(a) 3D and projection profiles;
(b) wave propagation along the $x$ axis at $t=0$ and $t=\pm5$.}
\vspace{0.2cm}

The third-order $2$-soliton solution in Figs. 14 and 15 displays the interaction between a second-order $1$-soliton and a simple $1$-soliton solutions. We can obviously find that the dynamical behavior of the high-order $N$-solitons are richer and more complicated than the one of the fundamental $N$-soliton.

\section{Conclusions}
In this paper, we have investigated the ngSS equation \eqref{ngss} through the assistance of the improved RHM. Starting from the temporal part of the Lax pair, we present two matrix functions and give their asymptotic properties. Then considering the spacial evolution, we successfully construct a suitable RHP. This method provides a pathway for addressing the symmetries inherent in the scattering data of Eq. \eqref{ngss}. Further on, under the reflectionless condition, the compact form of $N$-soliton solution is deduced. What's more, for the case of $1$-soliton, we theoretically analyze three cases. The general expressions of the amplitudes before and after the collision are addressed with the form of proposition.
For the purpose of deriving the $N_{0}$-th high-order $N$-soliton solution with the compact form, the perturbed terms and limiting techniques are implemented to the ngSS equation. Especially, when $N_{0}=1$, the $N_{0}$-th high-order $N$-soliton solution can be reduced to the general $N$-soliton solution.
It is obviously found that the high-order $N$-solitons of the ngSS equation \eqref{ngss} could have more complicated wave and trajectory structures, in addition, the dynamic behaviors of these figures are distinctly different from the simple $N$-soliton. Finally, intriguing graphical representations highlight novel characteristics inherent in these solutions. These solutions play a crucial role in revealing the abundant dynamics of solitons and advancing our comprehension of nonlocal nonlinear phenomena.

With the development of nonlocal integrable systems in fields of mathematics and physics, it is worthwhile to explore the possibility of applying these approaches to analyze other physically meaningful nonlocal nonlinear evolution equations. However, further examination and discussions on this matter will be reserved for future considerations. \\

\noindent
\textbf{Acknowledgments:} This work of the first author was supported by the National Natural Science Foundation of China (No.12271129) and the China Scholarship Council (No.202206
120152). The work of the second author was supported by the National Natural Science Foundation of China (No.12201622). The work of the third author was supported by the China Postdoctoral Science Foundation (2023M733404), the Young Innovative Talents Project of Guangdong Province of China (2022KQNCX104) and the Guangdong Basic and Applied Basic Research Foundation (2022A1515111209). The work of the fourth author was supported by the National Natural Science Foundation of China (No.12271129). The first author appreciates the hospitality of the Department of Mathematics, National University of Singapore, where the work was done.







\begin{thebibliography}{50}

\bibitem{Ablowitz-1974}
M.J. Ablowitz, D.J. Kaup, A.C. Newell and H. Segur. Inverse scattering
transform: Fourier analysis for nonlinear problems. \emph{Stud. Appl. Math.}, 53: 249-315, 1974.
\bibitem{Ablowitz-2013}
M.J. Ablowitz and Z.H. Musslimani. Integrable nonlocal nonlinear schr\"{o}dinger equation. \emph{Phys. Rev. Lett.}, 110: 064105, 2013.
\bibitem{Ablowitz-2016}
M.J. Ablowitz and Z.H. Musslimani. Inverse scattering transform for the
integrable nonlocal nonlinear Schr\"{o}dinger equation. \emph{Nonlinearity}, 29: 915-946, 2016.
\bibitem{Ablowitz-2018}
M.J. Ablowitz, X.D. Luo and Z.H. Musslimani. Inverse scattering transform for the nonlocal nonlinear Schr\"{o}dinger equation with nonzero boundary conditions. \emph{J. Math. Phys.}, 59: 011501, 2018.
\bibitem{SS-1991}
N. Sasa and J. Satsuma. New-type of soliton solutions for a higher-order nonlinear Schr\"{o}dinger equation. \emph{J. Phys. Soc. Jpn.}, 60: 409-417, 1991.
\bibitem{K-1985}
Y. Kodama. Optical solitons in a monomode fiber. \emph{J. Stat. Phys.}, 39: 597-614, 1985.
\bibitem{Xu-2013}
T. Xu and X.M. Xu. Single-and double-hump femtosecond vector solitons in the coupled Sasa-Satsuma system. \emph{Phys. Rev. E}, 87: 032913, 2013.
\bibitem{Xu-2015}
T. Xu, M. Li and L. Li. Anti-dark and Mexican-hat solitons in the Sasa-Satsuma equation on the continuous wave background. \emph{Europhys. Lett.}, 109: 30006, 2015.
\bibitem{J-2015}
J.J.C. Nimmo and H. Yilmaz. Binary Darboux transformation for the Sasa-Satsuma equation. \emph{J. Phys. A: Math. Theor.}, 48: 425202, 2015.
\bibitem{X-2014}
X. L\"{u}. Bright-soliton collisions with shape change by intensity redistribution for the coupled Sasa-Satsuma system in the optical fiber communications. \emph{Commun. Nonlinear Sci. Numer. Simul.}, 19: 3969-3987, 2014.
\bibitem{Zhao-2014}
L.C. Zhao, Z.Y. Yang and L. Ling. Localized waves on continuous wave background in a two-mode nonlinear fiber with high-order effects. \emph{J. Phys. Soc. Jpn.}, 83: 104401, 2014.
\bibitem{AM-2021}
A.M. Wazwaz and M. Mehanna. Higher-order Sasa-Satsuma equation: Bright and dark optical solitons. \emph{Optik}, 243: 167421, 2021.
\bibitem{Lv-2009}
C.-C. L\"{u} and Y. Chen. Symmetry and exact solutions of (2+1)-dimensional generalized sasa-satsuma equation via a modified direct method. \emph{Communications in Theoretical Physics}, 51: 973-978, 2009. 
\bibitem{Geng-2016}
X.G. Geng and J.P. Wu. Riemann-Hilbert approach and N-soliton solutions for a generalized Sasa-Satsuma equation. \emph{Wave Motion}, 60: 62-72, 2016.
\bibitem{Geng-2017}
J.P. Wu and X.G. Geng. Inverse scattering transform of the coupled Sasa-Satsuma equation by Riemann-Hilbert approach. \emph{Commun. Theor. Phys.}, 67: 527-534, 2017.
\bibitem{Zhu-2017}
C.Q. Song, D.M. Xiao and Z.N. Zhu. Reverse space-time nonlocal Sasa-Satsuma equation and its solutions. \emph{J. Phys. Soc. Jpn.}, 86: 054001, 2017.
\bibitem{Chen-2022}
M.M. Wang and Y. Chen. Novel solitons and higher-order solitons for the nonlocal generalized Sasa-Satsuma equation of reverse-space-time type. \emph{Nonlinear Dyn.}, 110: 753-769, 2022.
\bibitem{Zhu-2023}
H.-Q. Sun and Z.N. Zhu. Darboux transformation and soliton solution of the nonlocal generalized Sasa-Satsuma equation. \emph{Mathematics}, 11: 865, 2023.
\bibitem{Wang-2022}
G.X. Wang, X.-B. Wang and B. Han. Inverse scattering of nonlocal Sasa-Satsuma equations and their multisoliton solutions. \emph{Eur. Phys. J. Plus}, 137: 404, 2022.
\bibitem{Ma-2023}
Y.Q. Liu, W.-X. Zhang and W.-X. Ma. Riemann-Hilbert problems and soliton solutions for a generalized coupled Sasa-Satsuma equation. \emph{Commun. Nonlinear Sci. Numer. Simul.}, 118: 107052, 2023.
\bibitem{XB-2021}
X.-B. Wang and B. Han. The nonlinear steepest descent approach for long time behavior of the two-component coupled Sasa-Satsuma equation with a $5\times5$ Lax Pair. \emph{Taiwan. J. Math.}, 25(2): 381-407, 2021.
\bibitem{XW-2023}
X.W. Yan and Y. Chen. Reverse-time type nonlocal Sasa-Satsuma equation and its soliton solutions. \emph{Commun. Theor. Phys.}, 75: 075005, 2023.
\bibitem{CS-1967}
C.S. Gardner, J.M. Greene, M.D. Kruskal and R.M. Miura. Method for solving the Korteweg-de Vries equation. \emph{Phys. Rev. Lett.}, 19: 1095, 1967.
\bibitem{VE-1974}
V.E. Zakharov and A.B. Shabat. A scheme for integrating the nonlinear equations of mathematical physics by the method of the inverse scattering problem. I. \emph{Funct. Anal. Appl.}, 8: 226-235, 1974.
\bibitem{VE-1979}
V.E. Zakharov and A.B. Shabat. Integration of nonlinear equations of mathematical physics by the method of inverse scattering. II. \emph{Funct. Anal. Appl.}, 13: 166-174, 1979.

\bibitem{Biondini-2014}
G. Biondini and G. Kova$\check{\mathrm{c}}$i$\check{\mathrm{c}}$. Inverse scattering transform for the focusing nonlinear Schr\"{o}dinger equation with nonzero boundary conditions. \emph{J. Math. Phys.}, 55: 031506, 2014.
\bibitem{Biondini-2016}
G. Biondini, D. Kraus and B. Prinari. The three-component defocusing nonlinear Schr\"{o}dinger equation with nonzero boundary conditions. \emph{Commun. Math. Phys.}, 348: 475-533, 2016.
\bibitem{Yang-2018}
J.K. Yang. Physically significant nonlocal nonlinear Schr\"{o}dinger equation and its soliton solutions. \emph{Phys. Rev. E.}, 98: 042202, 2018.
\bibitem{Yan-2020}
G.Q. Zhang and Z.Y. Yan. Inverse scattering transforms and soliton solutions of focusing and defocusing nonlocal mKdV equations with non-zero boundary conditions. \emph{Physica D}, 402:  132170, 2020.
\bibitem{Ma-2020}
W.-X. Ma, Y.H. Huang and F.D. Wang. Inverse scattering transforms and soliton solutions of nonlocal reverse-space nonlinear Schr\"{o}dinger hierarchies. \emph{Stud. Appl. Math.}, 145:  563-585, 2020.
\bibitem{Ma-2022}
W.-X. Ma. Riemann-hilbert problems and inverse scattering of nonlocal real reverse-spacetime matrix akns hierarchies. \emph{Physica D}, 430: 430, 2022.
\bibitem{Feng-2021}
Y. Chen, B.-F. Feng and L.M. Ling. The robust inverse scattering method
for focusing Ablowitz-Ladik equation on the non-vanishing background. \emph{Physica D},  424: 132954, 2021.
\bibitem{GX-2023}
G.X. Wang and B. Han. The discrete modified Korteweg-de Vries equation under nonzero
boundary conditions. \emph{Appl. Math. Lett.}, 140: 108562, 2023.

\bibitem{Wu-2019}
J.P. Wu. Riemann-Hilbert approach of the Newell-type long-wave-short-wave equation via the temporal-part spectral analysis. \emph{Nonlinear Dyn.}, 98: 749, 2019.
\bibitem{Wu-2023}
J.P. Wu. A novel Riemann-Hilbert approach via t-part spectral analysis for a physically significant nonlocal integrable nonlinear Schr\"{o}dinger equation. \emph{Nonlinearity}, 36: 2021-2037, 2023.

\bibitem{Gagnon-1994}
L. Gagnon, N. Stivenart. N-soliton interaction in optical fibers: the multiple-pole case. \emph{Opt. Lett.}, 19: 619-621, 1994.
\bibitem{Villarroel-1999}
J. Villarroel, M.J. Ablowitz. A novel class of solutions of the non-stationary Schr\"{o}dinger and the Kadomtsev-Petviashvili I equations. \emph{Commun. Math. Phys.}, 207: 1-42, 1999.
\bibitem{Ablowitz-2000}
M.J. Ablowitz, S. Charkravarty, A.D. Trubatch, J. Villarroel. On the discrete spectrum of the nonstationary Schr\"{o}dinger equation and multipole lumps of the Kadomtsev-Petviashvili I equation. \emph{Phys. Lett. A}, 267: 132-146, 2000.
\bibitem{Bian-2015}
B. Bian, B.L. Guo, L.M. Ling. High-order soliton solution of Landau-Lifshitz equation. \emph{Stud. Appl. Math.}, 134: 181-214, 2015.
\bibitem{Bo-2019}
B. Yang and Y. Chen. High-order soliton matrices for Sasa-Satsuma equation via local
Riemann-Hilbert problem. \emph{Nonlinear Anal. Real.}, 45: 918-941, 2019.
\bibitem{Fan-2020}
Z.C. Zhang and E.G. Fan. Inverse scattering transform for the Gerdjikov-Ivanov equation with nonzero boundary conditions. \emph{Z. Angew. Math. Phys.}, 71: 149, 2020.
\bibitem{Tian-2022}
J.-J. Yang, S.-F. Tian and Z.-Q. Li. Riemann-Hilbert problem for the focusing nonlinear Schrdinger equation with multiple high-order poles under nonzero boundary conditions. \emph{Physica D}, 432: 133162, 2022.
\bibitem{Mao-2023}
J.J. Mao,T.Z. Xu and L.F. Shi. Soliton and breather solutions of the higher-order modified Korteweg-de Vries equation with constants background. \emph{Z. Angew. Math. Phys.}, 74: 78, 2023.
\bibitem{Bo-2018}
B. Yang and Y. Chen. Dynamics of high-order solitons in the nonlocal nonlinear
Schr\"{o}dinger equations. \emph{Nonlinear Dyn.}, 94: 489-502, 2018.
\bibitem{Min-2023}
M.M. Wang and Y. Chen. General multi-soliton and higher-order soliton solutions
for a novel nonlocal Lakshmanan-Porsezian-Daniel equation. \emph{Nonlinear Dyn.}, 111: 655-669, 2023.
\bibitem{Voros-1989}
A. Voros. Wentzel-Kramers-Brillouin method in the Bargmann representation. \emph{Phys. Rev. A}, 40(12): 6814, 1989.



\end{thebibliography}
\end{document}